\documentclass[preprint,showpacs,preprintnumbers,superscriptaddress,amsmath,amssymb]{revtex4}

\usepackage{graphicx}% Include figure files
\usepackage{dcolumn}% Align table columns on decimal point
\usepackage{bm}% bold math

\input psfig.sty
\def\figurewidth{\textwidth}
\def\figurewidthb{0.3\textwidth}
\def\figurewidthc{0.68\textwidth}

\def\bjet{b{\,\rm jet}}

\def\bjets{b{\,\rm jets}}
\def\cjet{c{\,\rm jet}}
\def\cjets{c{\,\rm jets}}
\def\ppbar{$p\overline{p}~$}             %ppbar
\def\met{\mbox{${\hbox{$E$\kern-0.6em\lower-.1ex\hbox{/}}}_T$}} %missing ET
\def\mex{\mbox{${\hbox{$E$\kern-0.6em\lower-.1ex\hbox{/}}}_x$}} %missing Ex
\def\mey{\mbox{${\hbox{$E$\kern-0.6em\lower-.1ex\hbox{/}}}_y$}} %missing Ey
\def\mexy{\mbox{${\hbox{$E$\kern-0.6em\lower-.1ex\hbox{/}}}_{x,y}$}} %missing Exy
 
\newcommand{\lapprox}{\stackrel{<}{_{\sim}}}

\begin{document}
%
% ======> Title of the paper goes here <====================
%
\title{\boldmath  Measurement of Cross Sections for $b$ Jet Production in Events with a $Z$ Boson
in \ppbar Collisions at $\sqrt{s}=1.96$~TeV}

\affiliation{Institute of Physics, Academia Sinica, Taipei, Taiwan 11529, Republic of China} 
\affiliation{Argonne National Laboratory, Argonne, Illinois 60439} 
\affiliation{University of Athens, 157 71 Athens, Greece} 
\affiliation{Institut de Fisica d'Altes Energies, Universitat Autonoma de Barcelona, E-08193, Bellaterra (Barcelona), Spain} 
\affiliation{Baylor University, Waco, Texas  76798} 
\affiliation{Istituto Nazionale di Fisica Nucleare Bologna, $^v$University of Bologna, I-40127 Bologna, Italy} 
\affiliation{Brandeis University, Waltham, Massachusetts 02254} 
\affiliation{University of California, Davis, Davis, California  95616} 
\affiliation{University of California, Los Angeles, Los Angeles, California  90024} 
\affiliation{University of California, San Diego, La Jolla, California  92093} 
\affiliation{University of California, Santa Barbara, Santa Barbara, California 93106} 
\affiliation{Instituto de Fisica de Cantabria, CSIC-University of Cantabria, 39005 Santander, Spain} 
\affiliation{Carnegie Mellon University, Pittsburgh, PA  15213} 
\affiliation{Enrico Fermi Institute, University of Chicago, Chicago, Illinois 60637} 
\affiliation{Comenius University, 842 48 Bratislava, Slovakia; Institute of Experimental Physics, 040 01 Kosice, Slovakia} 
\affiliation{Joint Institute for Nuclear Research, RU-141980 Dubna, Russia} 
\affiliation{Duke University, Durham, North Carolina  27708} 
\affiliation{Fermi National Accelerator Laboratory, Batavia, Illinois 60510} 
\affiliation{University of Florida, Gainesville, Florida  32611} 
\affiliation{Laboratori Nazionali di Frascati, Istituto Nazionale di Fisica Nucleare, I-00044 Frascati, Italy} 
\affiliation{University of Geneva, CH-1211 Geneva 4, Switzerland} 
\affiliation{Glasgow University, Glasgow G12 8QQ, United Kingdom} 
\affiliation{Harvard University, Cambridge, Massachusetts 02138} 
\affiliation{Division of High Energy Physics, Department of Physics, University of Helsinki and Helsinki Institute of Physics, FIN-00014, Helsinki, Finland} 
\affiliation{University of Illinois, Urbana, Illinois 61801} 
\affiliation{The Johns Hopkins University, Baltimore, Maryland 21218} 
\affiliation{Institut f\"{u}r Experimentelle Kernphysik, Universit\"{a}t Karlsruhe, 76128 Karlsruhe, Germany} 
\affiliation{Center for High Energy Physics: Kyungpook National University, Daegu 702-701, Korea; Seoul National University, Seoul 151-742, Korea; Sungkyunkwan University, Suwon 440-746, Korea; Korea Institute of Science and Technology Information, Daejeon, 305-806, Korea; Chonnam National University, Gwangju, 500-757, Korea} 
\affiliation{Ernest Orlando Lawrence Berkeley National Laboratory, Berkeley, California 94720} 
\affiliation{University of Liverpool, Liverpool L69 7ZE, United Kingdom} 
\affiliation{University College London, London WC1E 6BT, United Kingdom} 
\affiliation{Centro de Investigaciones Energeticas Medioambientales y Tecnologicas, E-28040 Madrid, Spain} 
\affiliation{Massachusetts Institute of Technology, Cambridge, Massachusetts  02139} 
\affiliation{Institute of Particle Physics: McGill University, Montr\'{e}al, Qu\'{e}bec, Canada H3A~2T8; Simon Fraser University, Burnaby, British Columbia, Canada V5A~1S6; University of Toronto, Toronto, Ontario, Canada M5S~1A7; and TRIUMF, Vancouver, British Columbia, Canada V6T~2A3} 
\affiliation{University of Michigan, Ann Arbor, Michigan 48109} 
\affiliation{Michigan State University, East Lansing, Michigan  48824}
\affiliation{Institution for Theoretical and Experimental Physics, ITEP, Moscow 117259, Russia} 
\affiliation{University of New Mexico, Albuquerque, New Mexico 87131} 
\affiliation{Northwestern University, Evanston, Illinois  60208} 
\affiliation{The Ohio State University, Columbus, Ohio  43210} 
\affiliation{Okayama University, Okayama 700-8530, Japan} 
\affiliation{Osaka City University, Osaka 588, Japan} 
\affiliation{University of Oxford, Oxford OX1 3RH, United Kingdom} 
\affiliation{Istituto Nazionale di Fisica Nucleare, Sezione di Padova-Trento, $^w$University of Padova, I-35131 Padova, Italy} 
\affiliation{LPNHE, Universite Pierre et Marie Curie/IN2P3-CNRS, UMR7585, Paris, F-75252 France} 
\affiliation{University of Pennsylvania, Philadelphia, Pennsylvania 19104}
\affiliation{Istituto Nazionale di Fisica Nucleare Pisa, $^x$University of Pisa, $^y$University of Siena and $^z$Scuola Normale Superiore, I-56127 Pisa, Italy} 
\affiliation{University of Pittsburgh, Pittsburgh, Pennsylvania 15260} 
\affiliation{Purdue University, West Lafayette, Indiana 47907} 
\affiliation{University of Rochester, Rochester, New York 14627} 
\affiliation{The Rockefeller University, New York, New York 10021} 
\affiliation{Istituto Nazionale di Fisica Nucleare, Sezione di Roma 1, $^{aa}$Sapienza Universit\`{a} di Roma, I-00185 Roma, Italy} 

\affiliation{Rutgers University, Piscataway, New Jersey 08855} 
\affiliation{Texas A\&M University, College Station, Texas 77843} 
\affiliation{Istituto Nazionale di Fisica Nucleare Trieste/Udine, $^{bb}$University of Trieste/Udine, Italy} 
\affiliation{University of Tsukuba, Tsukuba, Ibaraki 305, Japan} 
\affiliation{Tufts University, Medford, Massachusetts 02155} 
\affiliation{Waseda University, Tokyo 169, Japan} 
\affiliation{Wayne State University, Detroit, Michigan  48201} 
\affiliation{University of Wisconsin, Madison, Wisconsin 53706} 
\affiliation{Yale University, New Haven, Connecticut 06520} 
\author{T.~Aaltonen}
\affiliation{Division of High Energy Physics, Department of Physics, University of Helsinki and Helsinki Institute of Physics, FIN-00014, Helsinki, Finland}
\author{J.~Adelman}
\affiliation{Enrico Fermi Institute, University of Chicago, Chicago, Illinois 60637}
\author{T.~Akimoto}
\affiliation{University of Tsukuba, Tsukuba, Ibaraki 305, Japan}
\author{B.~\'{A}lvarez~Gonz\'{a}lez}
\affiliation{Instituto de Fisica de Cantabria, CSIC-University of Cantabria, 39005 Santander, Spain}
\author{S.~Amerio$^w$}
\affiliation{Istituto Nazionale di Fisica Nucleare, Sezione di Padova-Trento, $^w$University of Padova, I-35131 Padova, Italy} 

\author{D.~Amidei}
\affiliation{University of Michigan, Ann Arbor, Michigan 48109}
\author{A.~Anastassov}
\affiliation{Northwestern University, Evanston, Illinois  60208}
\author{A.~Annovi}
\affiliation{Laboratori Nazionali di Frascati, Istituto Nazionale di Fisica Nucleare, I-00044 Frascati, Italy}
\author{J.~Antos}
\affiliation{Comenius University, 842 48 Bratislava, Slovakia; Institute of Experimental Physics, 040 01 Kosice, Slovakia}
\author{G.~Apollinari}
\affiliation{Fermi National Accelerator Laboratory, Batavia, Illinois 60510}
% ...

\collaboration{CDF Collaboration\footnote{With visitors from $^a$University of Massachusetts Amherst, Amherst, Massachusetts 01003,
$^b$Universiteit Antwerpen, B-2610 Antwerp, Belgium, 
$^c$University of Bristol, Bristol BS8 1TL, United Kingdom,
$^d$Chinese Academy of Sciences, Beijing 100864, China, 
$^e$Istituto Nazionale di Fisica Nucleare, Sezione di Cagliari, 09042 Monserrato (Cagliari), Italy,
$^f$University of California Irvine, Irvine, CA  92697, 
$^g$University of California Santa Cruz, Santa Cruz, CA  95064, 
$^h$Cornell University, Ithaca, NY  14853, 
$^i$University of Cyprus, Nicosia CY-1678, Cyprus, 
$^j$University College Dublin, Dublin 4, Ireland,
$^k$Royal Society of Edinburgh/Scottish Executive Support Research Fellow,
$^l$University of Edinburgh, Edinburgh EH9 3JZ, United Kingdom, 
$^m$Universidad Iberoamericana, Mexico D.F., Mexico,
$^n$Queen Mary, University of London, London, E1 4NS, England,
$^o$University of Manchester, Manchester M13 9PL, England, 
$^p$Nagasaki Institute of Applied Science, Nagasaki, Japan, 
$^q$University of Notre Dame, Notre Dame, IN 46556,
$^r$University de Oviedo, E-33007 Oviedo, Spain, 
$^s$Texas Tech University, Lubbock, TX  79409, 
$^t$IFIC(CSIC-Universitat de Valencia), 46071 Valencia, Spain,
$^u$University of Virginia, Charlottesville, VA  22904,
$^{cc}$On leave from J.~Stefan Institute, Ljubljana, Slovenia, 
}}
\noaffiliation

%\date{\today}

%
% ==============> Text of the abstract goes here <=====================
% 
\begin{abstract}
A measurement of the $\bjet$ production cross section is presented
for events containing a $Z$ boson produced in $p\bar{p}$ collisions
at $\sqrt{s}=1.96$~TeV, using data corresponding to an integrated luminosity
of $2$~fb$^{-1}$  collected by the CDF II
detector
at the Tevatron. $Z$ bosons are selected in the electron and muon decay
modes. Jets are considered with transverse energy $E_T>20$~GeV 
and pseudorapidity
$|\eta|<1.5$ and are identified as $\bjets$
using a secondary vertex algorithm. The 
ratio of the integrated
$Z+\bjet$ cross section to the inclusive $Z$ production cross section 
is measured to be $3.32 \pm 0.53 {\rm (stat.)} \pm 0.42  {\rm (syst.)}\times 10^{-3}$. 
This ratio is also measured differentially
in jet $E_T$, jet $\eta$, $Z$-boson transverse momentum, number of jets, and number of $\bjets$.
The predictions from leading order Monte Carlo generators and next-to-leading-order
QCD calculations are found to be consistent with the measurements
within experimental and theoretical uncertainties.

\end{abstract}

\maketitle

%\runninglinenumbers  % LINENO CTOBS

\section{Introduction}
The associated production of $Z$ bosons
and one or more $\bjets$ provides an important test of quantum-chromodynamics (QCD)
calculations,  for which the theoretical predictions for this process
vary significantly~\cite{zbmcfm,pythia,alpgen}. 
The understanding of this process and its description by 
current theoretical calculations is important since it
is the largest background, e.g., to the search for the standard model Higgs boson 
in the $ZH \to Zb\bar{b}$ decay mode~\cite{smhiggs} and to searches
for the supersymmetric partners of $b$ quarks~\cite{sbottomd0,sbottomcdf}.
The process is also sensitive to the $b$ quark density in the proton. 
A precise knowledge of the  $b$ quark density is necessary to accurately
predict processes that strongly depend on it such as 
electroweak production of single top quarks~\cite{singletop} or the production of Higgs
bosons within certain supersymmetric models~\cite{bbh,mssmhiggs}. 

\begin{figure}%[!ht]
  \begin{center}
% uncomment for prd
   \includegraphics[width=\figurewidthb]{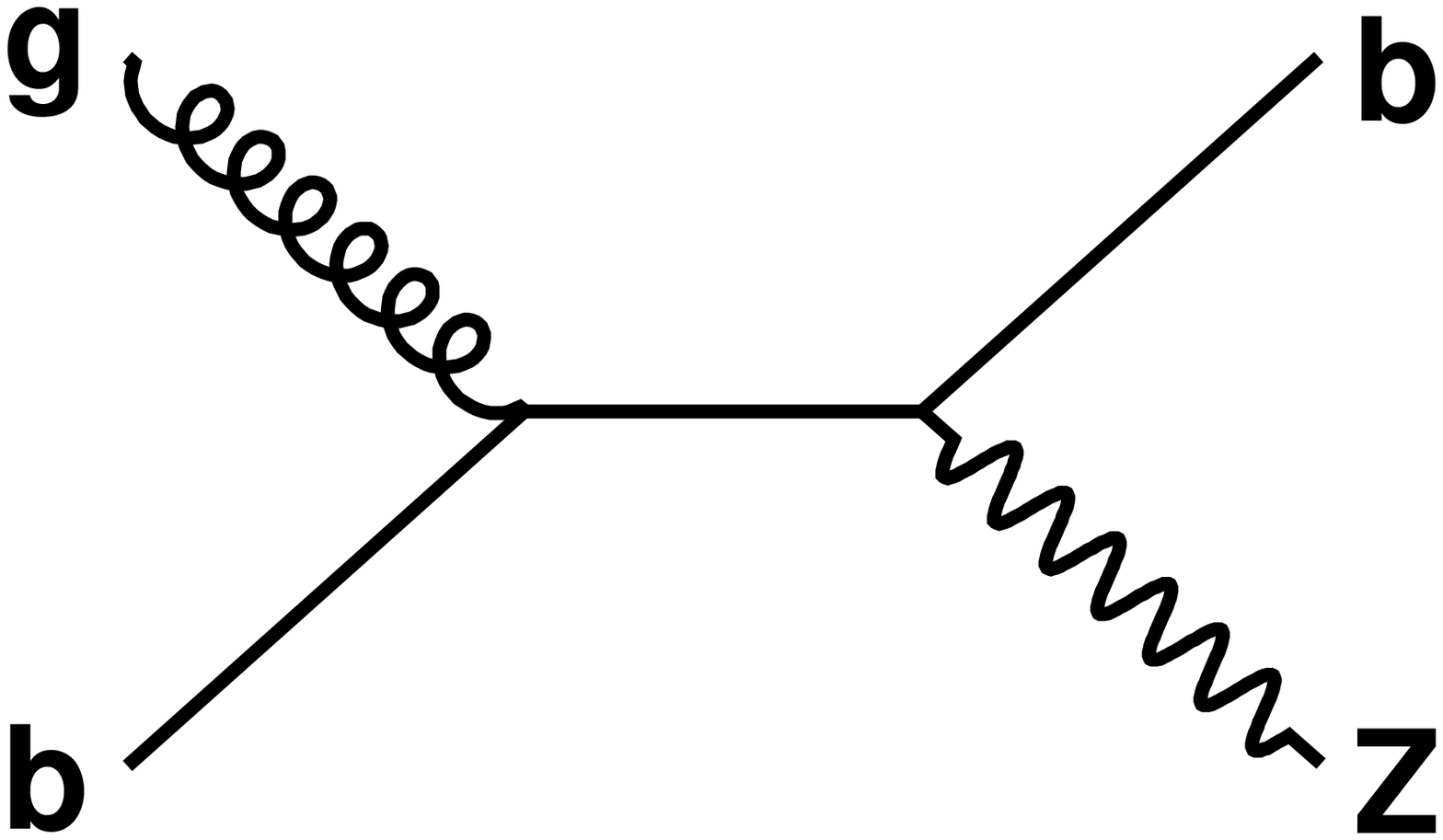}
   \includegraphics[width=\figurewidthb]{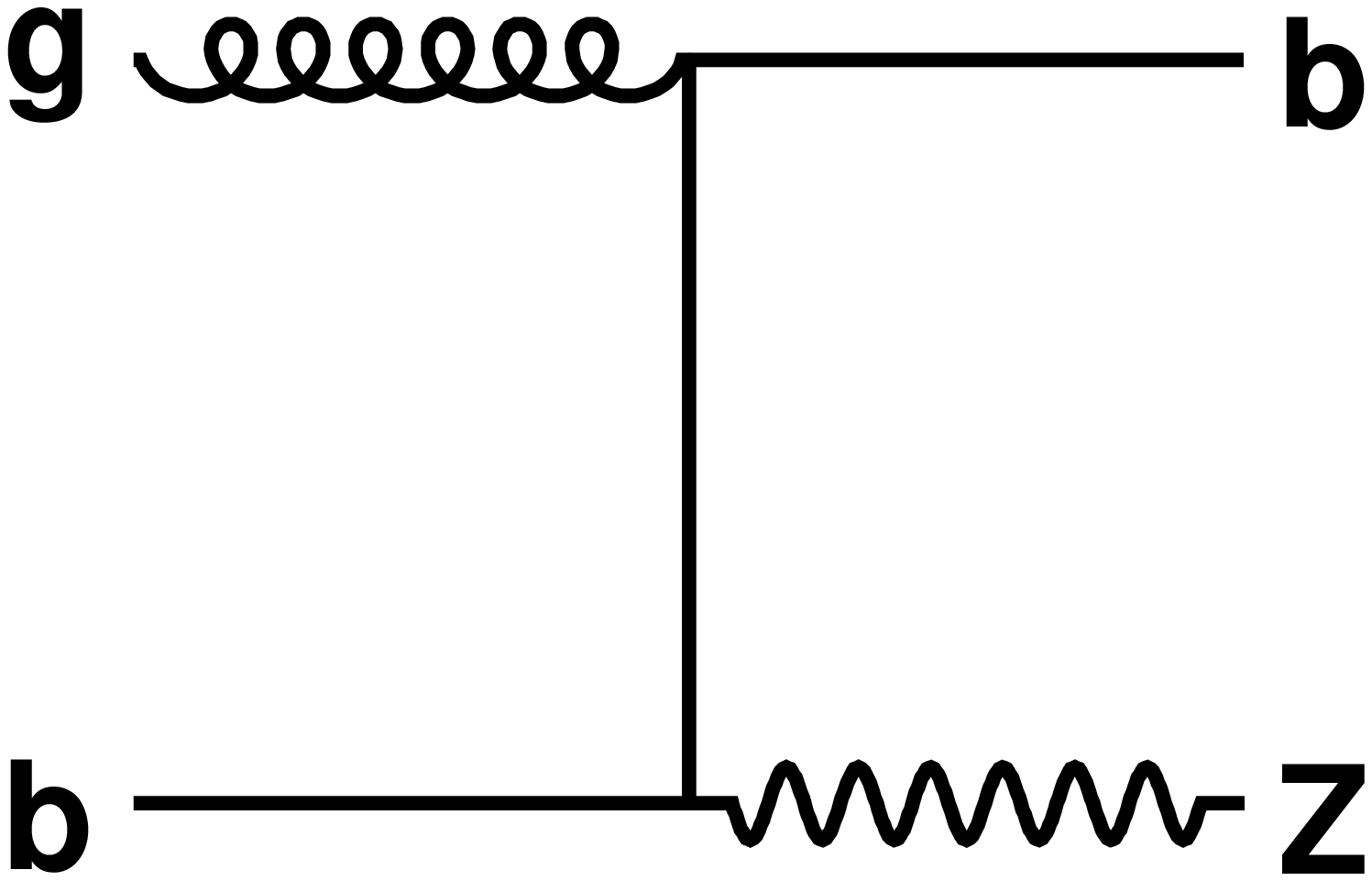}
   \includegraphics[width=\figurewidthb]{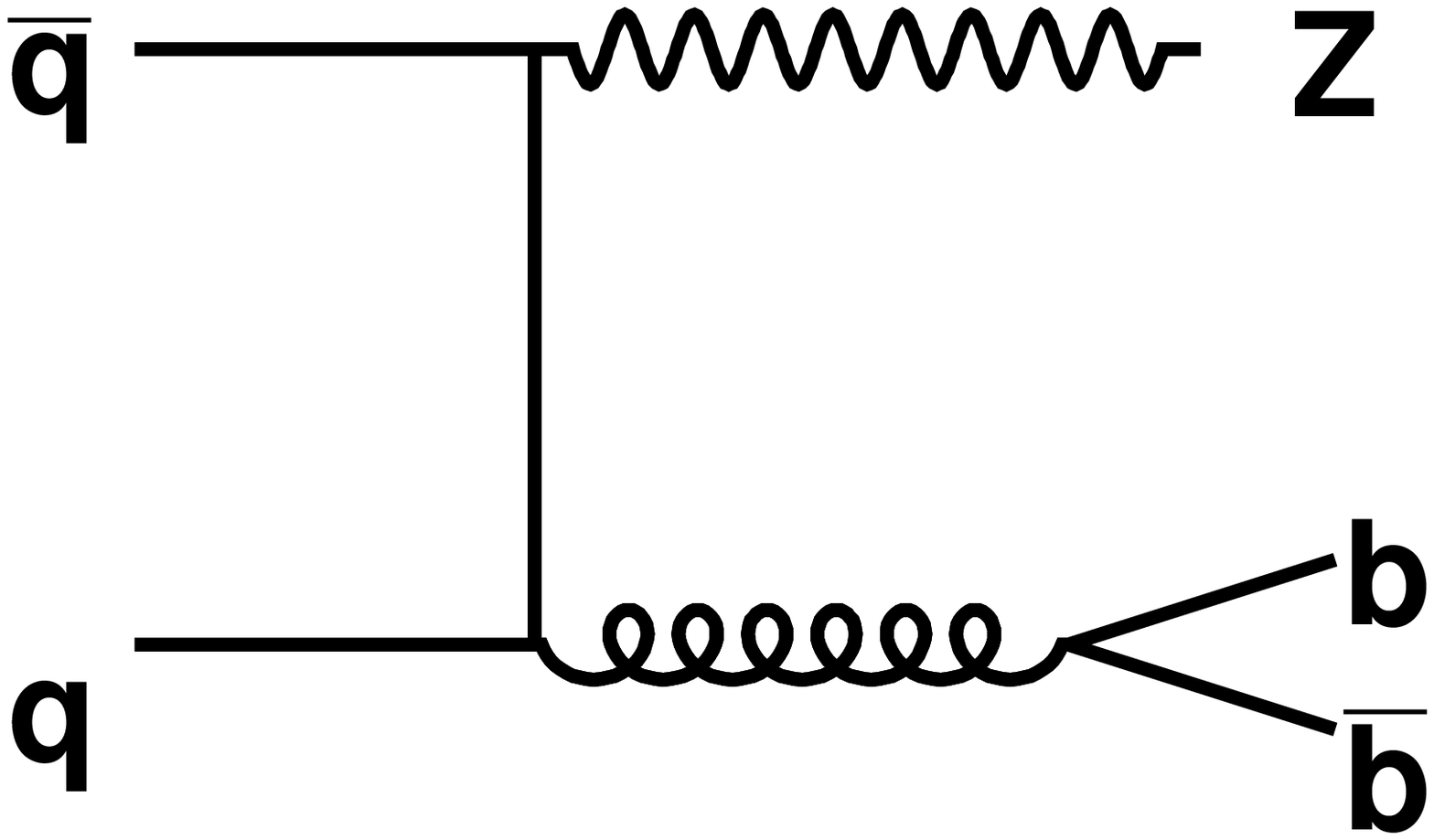}
  \end{center}
\caption{Leading order Feynman diagrams for $gb \to Zb$
and $q\bar{q} \to Zb\bar{b}$ production.
\label{fig:feyn}}
\end{figure}

The Feynman diagrams of the contributing leading order processes $gb
\to Zb$ and $q\bar{q} \to Zb\bar{b}$ are shown in
Fig.~\ref{fig:feyn}. In the first two diagrams a $b$ quark from the
proton undergoes a hard scatter and a $\bar{b}$ quark typically
remains close to the parent proton and may not be detected. In the
third diagram the $b\bar{b}$ quark pair can be produced close to each
other and may sometimes be reconstructed in the same jet (referred to
as a ``$b\bar{b}$ jet''). According to QCD calculations the latter diagram
is predicted to account for approximately 50\% of $b$ jet production
in association with a $Z$ boson at the Tevatron~\cite{zbmcfm}.

Previously the integrated cross section for $Z+\bjet$ production 
has been measured with an uncertainty of 39\% by the CDF collaboration~\cite{cdfzb}. 
The D0 collaboration also measured this process assuming the ratio of the
$Z+\bjet$ to $Z+\cjet$ cross section from next-to-leading order (NLO) 
QCD calculations~\cite{d0zb}. The cross section of $Z$+jets production has also
been measured recently by both collaborations and found to agree well 
with QCD calculations~\cite{cdfzjet, d0zjet}. A preliminary measurement 
has been made of the related  $W+\bjet$ process\cite{Wb}.

In this article we present an update to the integrated $Z+\bjet$ cross section
measurement with a substantially reduced uncertainty and for the
first time differential cross section measurements. The measurement is
made by selecting pairs of electrons or muons (dileptons) with an invariant mass
consistent with the mass of the $Z$ boson, $M_Z$, and jets containing a displaced 
secondary vertex
consistent with the decay of a bottom hadron. Contributions
from the decay of known heavy particles (such as $Z$ or top quarks) to $b$ hadrons are not included
in our definition of the cross section and are subtracted from the data.

The light and charm jets (i.e., jets that do not contain a $b$ hadron) 
remaining after this selection are
discriminated from $\bjets$ using the invariant mass of all charged
particles associated to the secondary vertex, exploiting the large
mass of the $b$ quark compared to the other partons. Throughout this
article we use $Z$ to denote any dilepton events due to $Z$ or
$\gamma^*$ production with an invariant mass
$76<M_{ll}<106$~GeV/$c^2$, albeit the contribution of virtual photons
is predicted to be below 1\% of the  $Z$ production rate~\cite{wzinclprl}.

We use data collected by the CDF II detector
at the Tevatron $p\bar{p}$ collider between February 2002 and May 2007, corresponding
to an integrated luminosity of 2.0~fb$^{-1}$.

\section{The CDF~II Detector}
The CDF~II detector is described in detail elsewhere~\cite{CDF_run2}
and consists of a precision tracking system, electromagnetic and
hadronic calorimeters and muon spectrometers. The tracking detector is
coaxial with the beam-pipe and consists of silicon strip
detectors~\cite{silicon} surrounded by a wire drift chamber
(COT)~\cite{COT} inside a $1.4$~T magnetic field provided by a
solenoid. The solenoid is surrounded by electromagnetic~\cite{cem,pem}
and hadronic calorimeters~\cite{cha} that use lead and stainless steel
as absorber materials, respectively, and scintillators as active
material. Inside the electromagnetic calorimeter 
a proportional strip and wire chamber is embedded  at about 6 radiation
lengths, providing
an accurate position measurement~\cite{ces}.  The muon
detectors~\cite{cmu} surround the calorimeters and consist of
wire chambers and scintillators. Gas $\check{\textrm{C}}$erenkov
counters, located close to the beampipe, are used to measure a fraction of 
the inelastic event rate and thereby the collider luminosity~\cite{CLC}.

A cylindrical coordinate system is used in which the $z$ axis is along the 
proton beam direction and $\theta$ is the polar angle.
The pseudorapidity is defined as $\eta = -\ln \tan(\theta/2)$, while 
the transverse momentum is 
given by $p_T= p\sin\theta$ and the transverse energy by $E_T=E \sin\theta$.
Missing transverse energy, $\met$, is defined as the magnitude of $-\Sigma_{i}E^{i}_T\hat{n}_i$,
where $E_T^i$ is the transverse energy deposited in the $i^{th}$ calorimeter tower 
and $\hat{n}_i$ is a unit vector pointing from the beamline to the $i$th tower in the 
azimuthal plane.

\section{Measurement Procedure}
In this article we present a measurement of the ratio of the cross
section for $Z+\bjet$ production to the inclusive $Z$ production cross
section. Measuring the ratio has the advantage that several
uncertainties, e.g., on the integrated luminosity and on the lepton
identification, largely cancel. We present both per jet and per event
cross section ratios. The per jet cross section ratio is proportional
to the number of $\bjets$, while the per event
cross section ratio is proportional
to the number of events with one or more $\bjets$.
The per jet cross section
selection efficiency is independent of the number of $\bjets$ in the
event as it is proportional directly to the efficiency of identifying
a $\bjet$, $\epsilon_{\bjet}$, and thus has a smaller overall
error. For certain measurements, such as the cross section ratio as a
function of the number of jets, it is preferable to use an event based
definition. In this case the event selection efficiency depends on the
number of $\bjets$ in the event: for events with one $\bjet$ it is
also simply proportional to $\epsilon_{\bjet}$ while for events with
two $\bjets$ the efficiency for, e.g., identifying at least one
$\bjet$ is proportional to $\epsilon_{\bjet}\cdot(2-\epsilon_{\bjet})$.

The cross section  ratios are determined by
\begin{eqnarray}
\frac{\sigma^{\rm jet}(Z+\bjet)}{\sigma(Z)} & = &
\frac{N^{\rm jet}(Z+\bjet)/N(Z)}{\epsilon^{\rm jet}(Z+\bjet)/\epsilon(Z)}\\
\frac{\sigma^{\rm evt}(Z+N_{\bjet})}{\sigma(Z)} & = &
\frac{N^{\rm evt}(Z+N_{\bjet})/N(Z)}{\epsilon^{\rm evt}(Z+N_{\bjet})/\epsilon(Z)},
\end{eqnarray}
where $N(Z)$ is the number of events in the data with a $Z$ boson and
$\epsilon(Z)$ is the efficiency $\times$ acceptance for the $Z$ boson
selection within the $M_{ll}$ range of this analysis.  For the per jet
cross sections, $\sigma^{\rm jet}(Z+\bjet)$, the number of estimated
$\bjets$ in data for events with a $Z$ boson is $N^{\rm
jet}(Z+\bjet)$. For the per event cross sections, $\sigma^{\rm
evt}(Z+N_{\bjet})$, the number of data events with a $Z$ boson is
$N^{\rm evt}(Z+N_{\bjet})$, with each event having the number of
$\bjets$ equal to $N_{\bjet}$. All quantities are quoted after
background subtraction.  The quantities $\epsilon^{\rm jet}(Z+\bjet)$
and $\epsilon^{\rm evt}(Z+N_{\bjet})$ are the corresponding efficiency
$\times$ acceptances for the $Z+\bjet$ and the $Z+N_{\bjet}$
selections, respectively.

In the following, the selection and the methods for determining the efficiencies
and the number of $\bjets$ are described. The Monte Carlo 
simulation is tuned to reproduce the trigger, lepton, and $\bjet$ efficiencies
as measured in the data, and is used to correct the data for all detector
effects such as acceptance losses, efficiencies and resolutions. 

\section{Monte Carlo Simulation}
We use {\sc pythia}~\cite{pythia} and {\sc alpgen}~\cite{alpgen}
as the main Monte Carlo generators. For the {\sc pythia} generation the inclusive 
Drell-Yan process for 
$Z$ production is used, and jets are generated via the parton shower. 
This process includes matrix-element inspired corrections to the parton
shower to better describe the $p_T$ distribution of the $Z$ boson~\cite{pythiaz}. 
{\sc alpgen} calculates the leading order (LO) matrix elements separately for 
each parton emission and 
then matches to a parton shower from {\sc pythia}.
Double-counting between the matrix-element calculations and the parton-showers is avoided by using
the MLM matching procedure~\cite{mlm}.
For both {\sc pythia} and {\sc alpgen} 
the CTEQ5L~\cite{cteq5l} structure function is used for the parton distribution functions and 
``Tune A'' is used for the underlying event~\cite{ue,rickdwt}. 
For the modeling of $c$ and $\bjets$ a combination of {\sc pythia} and {\sc alpgen} is used:
the samples are averaged using equal portions of both since this 
gives the best description of the $E_T$ and $\eta$ 
distribution of the $\bjets$.
For the description of light jets we use only the {\sc pythia} sample which
gives a good description of inclusive $Z$+jet production at low jet multiplicities that 
are relevant for this analysis. The decays of $b$ hadrons are performed
using {\sc evtgen}~\cite{evtgen}.  The generated events are passed through the 
{\sc geant}3-based~\cite{geant} CDF detector simulation~\cite{cdfsim},
and thereafter reconstructed and analyzed in the same way as the data.

\section{Event Selection}
\label{sec:selection}
$Z$ boson candidates are identified in events with a dilepton pair which have an
invariant mass $M_{ll}$ between $76$ and $106$~GeV/$c^2$ where $\ell=e,\mu$. 

Events in the electron channel are triggered online by either one
central ($|\eta|<1.1$) electromagnetic calorimeter cluster with
$E_T>18$~GeV and a track with $p_T>9$~GeV/$c$ associated to it, or by
two electromagnetic clusters with $E_T>18$~GeV and $|\eta|<3.2$, where
no track association is required. These requirements are also made in the offline analysis.
Furthermore, all central electrons are required to have $E_T>10$~GeV and a 
matched track with $p_T>5$~GeV/$c$, and all forward electrons are required to have $E_T>18$~GeV. 
For forward electrons ($1.1<|\eta|<3.2$) a track requirement is not
imposed unless both electrons are forward, in which case they 
are required to have $E_T>25$~GeV and a matched track with $p_T>10$~GeV/$c$.
The electrons also have to pass certain quality criteria to verify
that they are consistent with the electromagnetic shower
characteristics  expected for electrons~\cite{wzinclprl}.

Events in the muon channel are triggered on at least one muon
candidate that has a signal in one of the muon chambers with
$|\eta|<1.0$ and $p_T>18$~GeV/$c$. In the offline analysis 
the second muon candidate is not
required to have a signal in the muon chambers but it must have hits
in the COT which reduces the acceptance to $|\eta| \lapprox 1.5$. All
muons are required to have calorimeter energy deposits consistent with
those expected from a minimum ionizing particle~\cite{wzinclprl}.

All leptons are required to be isolated from other particles in the event by a 
distance of $\Delta R=\sqrt{(\Delta\phi)^2+(\Delta \eta)^2}>0.4$, and 
at least one of the two muons (electrons) must have $p_T>18$~GeV/$c$ ($E_T>18$~GeV) 
while the second one is only required to have $p_T>10$~GeV/$c$ ($E_T>10$~GeV). 
However, for dielectron events where the two electrons are in the forward calorimeter, both are
required to have $E_T>18$~GeV to match the trigger requirements.

In order to reduce the background from particles that fake electrons or muons
the  two leptons in each event are required to have opposite charge. This cut
is not applied in the electron channel if one or both electrons are forward,
since the charge determination is not very precise in this region of the detector~\cite{wchargeasym}.

Using this selection we observe 193,749 $Z\to e^+e^-$ and 101,967
$Z\to \mu^+\mu^-$ candidates. The selection has $\epsilon(Z)=41\%$ for
$Z\to e^+e^-$ and $\epsilon(Z)=23\%$ for $Z\to \mu^+\mu^-$ events.

Jets are selected using a cone based algorithm with a cone size of $\Delta R=0.7$~\cite{cone}. 
This choice of cone size has the advantage over smaller cone sizes in
that the hadronization corrections are smaller.
The jets are measured 
in the calorimeter and corrected to the hadron level~\cite{jesnim}, i.e., they are corrected for the 
CDF calorimeter response and multiple $p\bar{p}$ interactions. Note that the jets are not 
corrected for the underlying event (underlying event correction) or
 any changes in the energy
contained within the jet cone due to fragmentation and any energy loss due to out-of-cone 
parton radiation (hadronization correction). In order 
to compare to parton level calculations, these additional corrections are determined and
applied to the theoretical calculation as described later. We observe 29,363 $Z\to e^+e^-$ 
and 18,087 $Z\to \mu^+\mu^-$ candidate events with at least one jet with $E_T>20$~GeV 
and $|\eta|<1.5$.

A $\bjet$ is defined at the hadron level as any jet that contains a
$b$ hadron within its cone. A secondary vertex algorithm is used
to identify $\bjets$ based on tracks with
$p_T>0.5$~GeV$/c$  that are displaced from the
primary vertex, exploiting the relatively long lifetime of $b$ hadrons,
as described in detail in Ref.~\cite{secvtx}.   Jets with a reconstructed secondary
vertex are denoted as ``tagged'' jets. The $b$ tagging efficiency
varies between 30\% and 40\% in the $E_T$ range relevant for this
analysis, and has been measured using data with an uncertainty of
$5.3\%$. The algorithm also tags about 8\% of $\cjets$ and 0.5\% of
light jets as determined with Monte Carlo simulation. 

A sign is assigned depending on whether the secondary vertex is in the
same hemisphere as the jet (positive tag) or in the opposite
hemisphere (negative tag). For $b$ jets the direction of the vertex
tag is aligned with the jet direction generally yielding a positive
tag, while for misreconstructed secondary vertices from light jets the
two directions are uncorrelated, yielding similar amounts of negative
and positive tags. Since the jets with negative tags are used in the fit
to determine the fraction of $b$ jets (see section~\ref{sec:bfrac}),
it is necessary to verify that the ratio of negatively to positively
tagged light jets in the Monte Carlo reproduces that in the data.  The
ratio has been measured in inclusive jet production as $0.65 \pm 0.07$,
in good agreement with the simulation value of $0.62$.

Events are rejected if 
$\met>25$~GeV and the sum of the transverse energies of all jets, leptons
and $\met$ is greater than $150$~GeV. 
These cuts reduce the background from $t\bar{t}$ production by a factor 10, while 
retaining 99\% of the signal.

The efficiency of this selection is $\epsilon^{\rm jet}(Z+\bjet)=8.7\%$.
In the data we observe 648 positively tagged jets and 151
negatively tagged jets. There are nine events that contain
two tagged jets. For these events all tags are found to be positive. 

The sample of tagged events contains a small
amount of background from known processes which have a true $b$ jet
and a larger background contribution from events where a $c$ jet or a light
jet has produced a secondary vertex tag. These backgrounds are
discussed in the next two sections.

\section{Background Contributions}
\label{sec:bgs}
The most important backgrounds that contain a true $b$ jet arise from
 $ZZ$ and $t\bar{t}$ production and from processes  where  one
or  two jets are misidentified as a lepton. This latter contribution arises
mainly from  $W+$jets events where one jet is misidentified 
 and multi-jet production where two jets are misidentified. Background
from non-$b$ jets is discussed in section~\ref{sec:bfrac}.

The $ZZ$ and $t\bar{t}$ backgrounds are determined using {\sc pythia}
Monte Carlo simulation.  The $ZZ$ Monte Carlo simulation is normalized
 to the  NLO QCD cross section calculation of $37.2$~fb~\cite{zz}, which is
the part of the cross section where both $Z$s are in
the mass range $76<M_{Z}<106$~GeV/$c^2$ and  where one $Z$ decays to any of
the charged leptons and the other to $b\bar{b}$.  The Monte Carlo simulation also
produces events for $\gamma^*/Z$ production outside this mass range
and for all standard model decays. The $t\bar{t}$ cross section is
taken from NLO QCD as $6.7$~pb~\cite{ttbar}. We estimate an
uncertainty on these backgrounds of 20\%, which takes into account the
uncertainty in the theoretical prediction for the production cross section and in the
experimental acceptance for our analysis. Backgrounds from $Z\to
\tau^+\tau^-$ and $WW$ production were also studied.  Both were found 
to be small, with $Z\to \tau^+\tau^-$ contributing 0.3 and $WW$
contributing $<0.01$ to the number of tagged jets.

The backgrounds due to jets being
misidentified as leptons are determined using the data. For the
dielectron channel a ``fake rate'' method is used where the fraction
of jets misidentified as electrons is measured in inclusive jet samples
and then applied to the jets in a sample of data events with one reconstructed
electron. This technique is
described in more detail in Refs.~\cite{cdfzb} and~\cite{trilprd}. The
uncertainty on this background is estimated at 50\%, using the
agreement between data and Monte Carlo simulation in the sidebands of
the $Z$ mass distribution. 
 For the dimuon channel we use events in
which both muons have the same electric charge since the chance of
faking a muon is assumed to be charge independent. The statistical error on this number
of events is used as the uncertainty.  The resulting background estimates for
the number of tagged jets (including both positive and negative tags) are shown in
Table~\ref{tab:bg}. 

\begin{table}[htbp]
\begin{center}
\caption{
Estimated numbers of background tagged jets (positive and negative)  
 for the $e^+e^-$ and the $\mu^+\mu^-$ channels.
The uncertainties include both statistical and systematic errors.\label{tab:bg}}
\vspace{0.1cm}
\begin{tabular}{lrr} 
\hline \hline
Background source & $e^+e^-$ & $\mu^+\mu^-$ \\\hline 
$ZZ$   & $6.5 \pm 1.3$ & $4.3 \pm 0.9$ \\
$t\bar{t}$   & $1.3 \pm 0.3$ & $1.4 \pm 0.3$ \\
$Z\rightarrow \tau^+\tau^-$ / $WW$   & $0.2 \pm 0.1$ & $0.1 \pm 0.1$ \\
fake lepton   & $16.4 \pm 8.2$ & $5.0 \pm 2.2$ \\\hline \hline
\end{tabular}
\end{center}
\end{table}

\begin{figure}%[!ht]
  \begin{center}
% uncomment for prd
    \includegraphics*[width=\figurewidth]{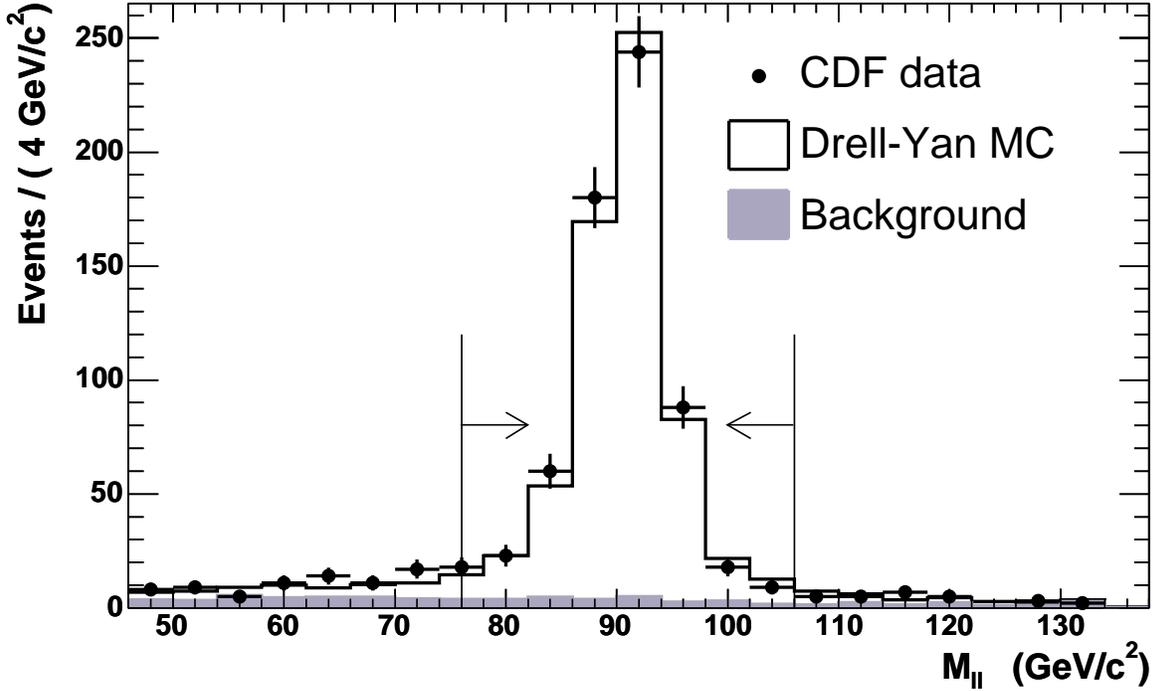}
  \end{center}
    \caption{Dilepton invariant mass for events with at least one positive secondary vertex tag. 
The data (points) are shown together with the Drell-Yan Monte Carlo (open histogram) and the
sum of the background contributions (filled histogram). The range where the data
are selected for this analysis is indicated by arrows.
\label{fig:zmass}}
\end{figure}

The dilepton invariant mass is shown in Fig.~\ref{fig:zmass} for events with at
least one positively tagged jet for the data, the 
{\sc pythia} signal Monte Carlo sample, and the background processes. 
The signal Drell-Yan Monte Carlo sample is normalized such that the  expectation equals the 
number of data events in  the range $76<M_{ll}<106$~GeV/$c^2$. The shape of 
expectation  agrees well with the data distribution
both in the peak, where the Drell-Yan signal dominates, and in the tails, where the background is significant.

\section{Determination of the Fraction of \boldmath $b$ Jets}
\label{sec:bfrac}
After the selection described in section~\ref{sec:selection} the sample
of tagged jets contains a significant fraction of light and charm
jets. Since the $Z+c$ jet cross section in the data is unknown and the
simulation may not accurately describe the rate of light jets that are
reconstructed with a secondary vertex, the fraction of $\bjets$ in the
data is determined using a likelihood fit of the invariant mass
distribution of the tracks forming the secondary vertex
$M_{SVTX}$. Due to the different masses of the quarks this distribution
enables a good discrimination between light, $c$, and $b$ jets.  It should
be noted that these distributions are affected, particularly for low values,
by the minimum track $p_T$ requirement 
and the efficiency of the algorithm to correctly
assign tracks to the secondary vertex.

The fit is performed for $M_{SVTX}<3.5$~GeV/$c^2$ where the data have
reasonable statistics. The $Z+$jets Monte Carlo simulation and the simulation
for the background processes, apart from fake leptons, are used to make
templates for the shape of the $M_{SVTX}$
distributions for light, $c$ and $\bjets$. 
The template for the shape of the background from
fake leptons is taken from the data. The normalization of each
background is fixed (see section~\ref{sec:bgs}), while the
normalizations of the light, $c$, and $b$ components of the $Z+$jets
are free parameters of the fit.

\begin{figure}%[h]
  \begin{center} 
  \includegraphics*[width=\figurewidthc]{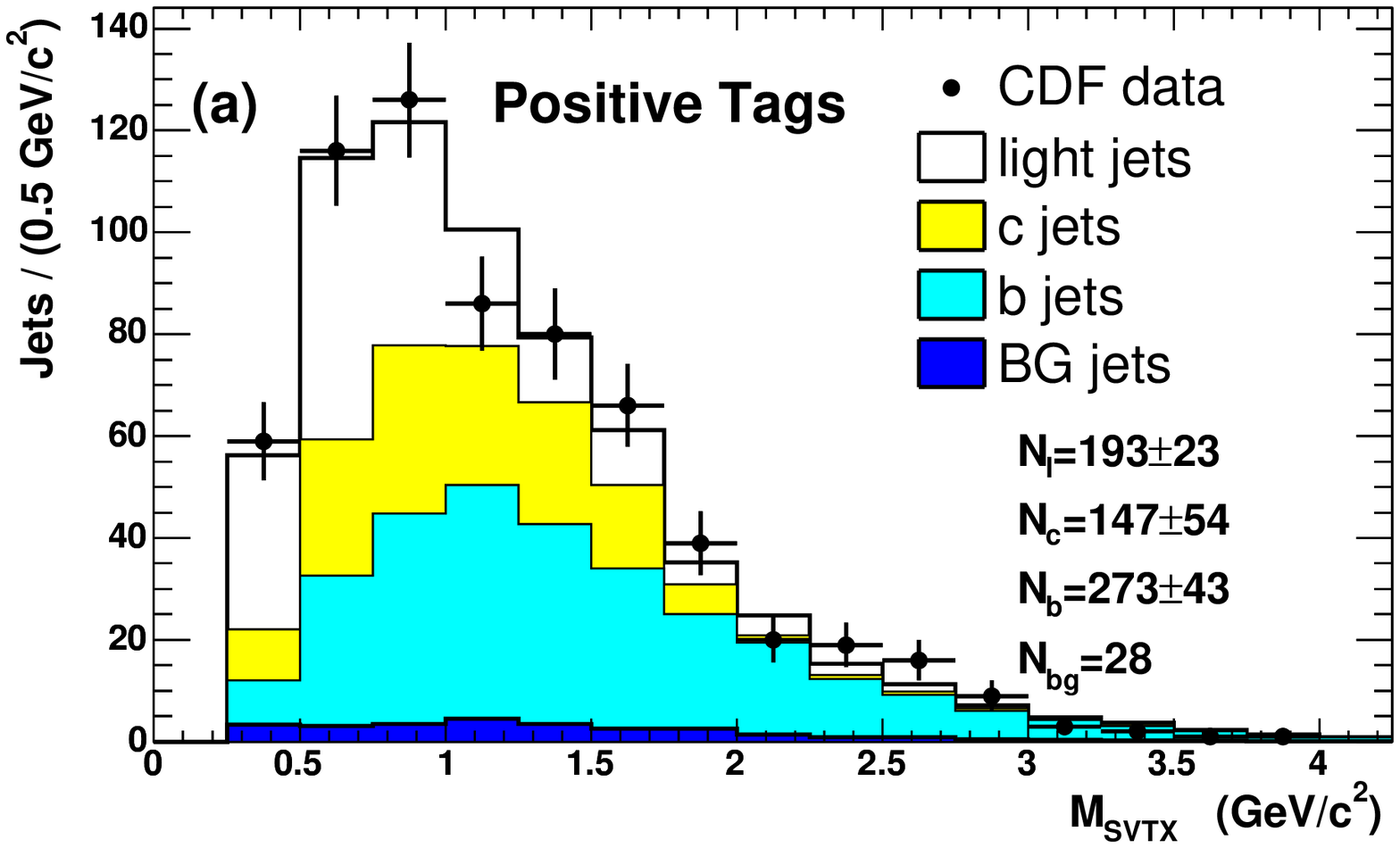}
   \includegraphics*[width=\figurewidthc]{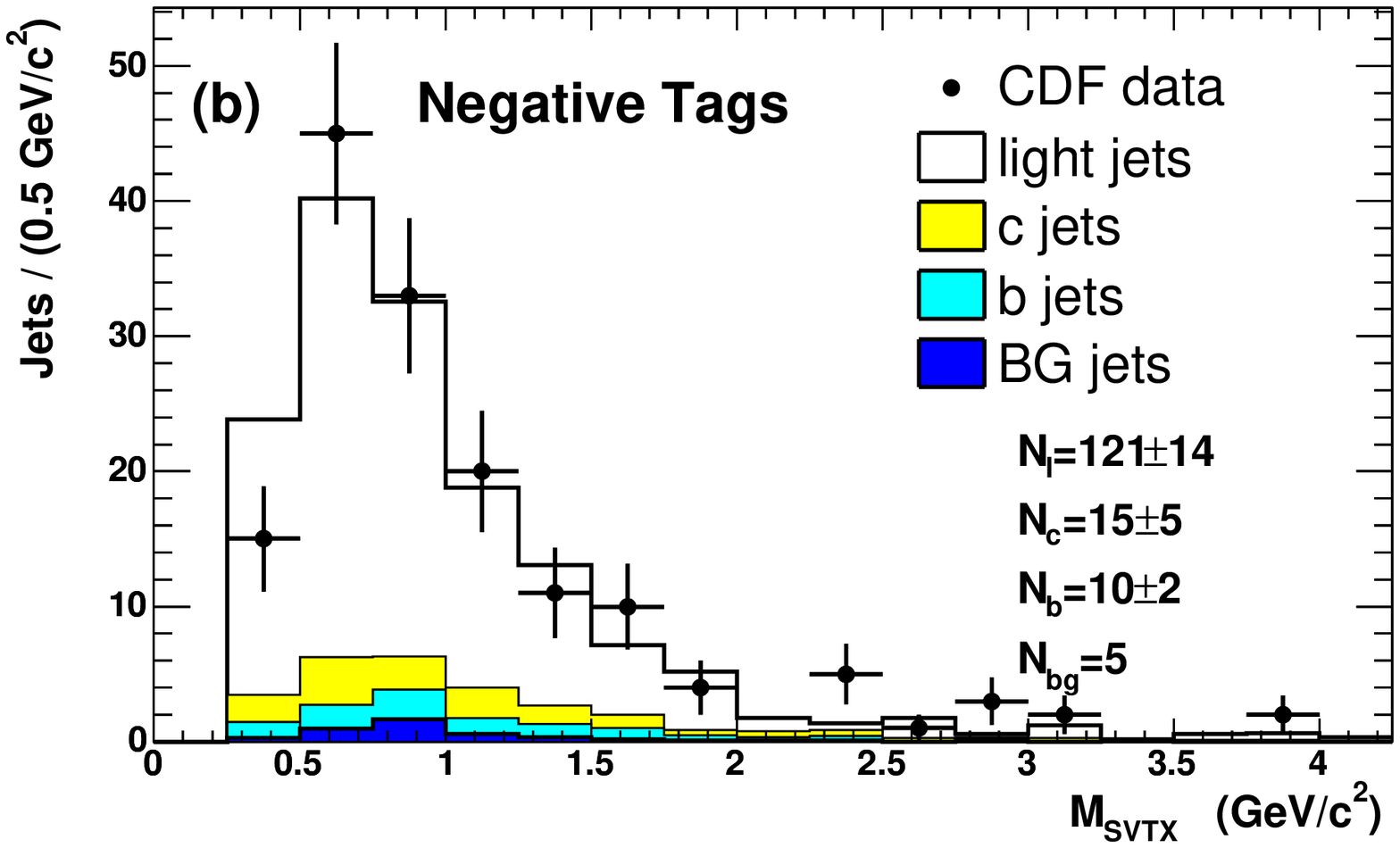}
  \end{center}
    \caption{Invariant mass of tracks at the secondary vertex for (a) positively and (b) negatively tagged jets.
Shown are the data (points) and the fitted contributions of light, $c$ and $\bjets$. 
Also shown is the background contribution from $Z$+light jets,  $Z+c$ jets, 
and from other processes with $b$ jets. The fitted number of light jets ($N_l$), $c$ jets ($N_c$),
$b$ jets ($N_b$), and the number of background events ($N_{bg}$) is also shown. \label{fig:msvtx}}
\end{figure}

This fit is done simultaneously for positively and negatively tagged
jets. We include the negatively tagged jets in the fit since this
results in a reduced uncertainty on the number of light and $c$ jets,
although the effect on the uncertainty for the number of $b$ jets is
marginal. The resulting fit is shown in Fig.~\ref{fig:msvtx}. It is
seen that for the positively tagged jets the $\bjets$ populate the
higher $M_{SVTX}$ values due to the large $b$ quark mass, allowing
them to be discriminated from the light and charm background that is
concentrated at low $M_{SVTX}$. The negatively tagged jets are mainly
populated by light jets, which are thus constrained by including this
distribution in the fit. The fit yields the number of positively tagged
$b$ jets as $N_b=270\pm 43$.
%The number of
%positively tagged light jets from the fit is $N_l=193\pm 22$ is
%above that predicted by  {\sc pythia} of 120. 
The correlation coefficient between the number of $b$ and $\cjets$ is
$-0.78$ and between the number of $\bjets$ and light jets is
$+0.22$. As a consistency check, when fitting only the positively
tagged jets, the result of $N_b=273\pm 44$ is consistent with the
default fit.

This technique for estimating
the number of $\bjets$ is used for the integrated and all differential cross section measurements 
except for the $Z+2\bjets$ measurement for which the statistics 
are too low to use this fit procedure. Since the background for double-tagged events from $c$ and 
light flavor jets is predicted from Monte Carlo to be only $0.79$ events,
compared to $9$ observed data events we simply subtract those components using 
the prediction.

\section{Systematic Uncertainties}
There are several sources of systematic uncertainty that are all 
evaluated separately for the integrated $Z+\bjet$ cross section and for each bin of the
differential measurements. 
Table~\ref{tab:syst} lists each source of systematic uncertainty 
and its effect on the integrated cross section ratio. 

\begin{table}[h]
\begin{center}
\caption{\label{tab:syst}
The systematic uncertainties on the measurement of the ratio 
${\sigma^{\rm jet}(Z+\bjet)}/{\sigma(Z)}$.
The total systematic uncertainty is estimated by adding the individual uncertainties in
quadrature.}

\begin{tabular}{lr} \hline \hline
Source of Uncertainty & Uncertainty (\%)\\ \hline
MC $E^{\rm jet}_T$ dependence    & 8.0 \\
MC $\eta^{\rm jet}$ dependence   & 2.8 \\
track finding efficiency  & 5.7 \\
$b$ quark fragmentation          & 0.8 \\
$b\bar{b}/b$, $c\bar{c}/c$ jet fractions   & 3.8 \\
light jet template               & 1.7 \\
$b$-tagging efficiency           & 5.3 \\
jet energy scale                 & 2.4 \\
misidentified lepton background  & 1.9 \\
other backgrounds                & 0.8 \\ \hline
total                            & 12.7 \\ \hline \hline
\end{tabular}
\end{center}
\end{table}

The largest systematic uncertainties arise from the Monte Carlo modeling of 
the $\bjet$ $E_T$ distribution and
the shape of the templates used for the extraction of the $\bjet$ fraction. 

The uncertainty on the $E_T$ dependence of 
$\bjets$ in $Z+\bjet$ production is estimated directly from the data by 
reweighting the shape of the simulated
distribution maintaining consistency with the data at the $1\sigma$
level. These variations are of a similar magnitude as the differences between the {\sc pythia}
and {\sc alpgen} generators. The resulting uncertainty  is 8.0\%. The same technique is used
to estimate a systematic uncertainty due to the modeling of the jet $\eta$ 
distribution of $2.8\%$.

Systematic uncertainties on the shape of the $M_{SVTX}$ templates are
estimated by varying the track finding efficiency by 3\%, by changing
the $b$ quark fragmentation function, and by varying the fraction of
$b\bar{b}$ to $\bjets$ (and $c\bar{c}$ to $\cjets$) between zero and
three times their default values in the simulation. These result in cross
section uncertainties of 5.7\%, 0.8\%, and 3.8\%, respectively. For the light jet
template the relative contribution of negative tags with respect to
positive tags is varied by 25\%, resulting in a cross section uncertainty of 1.7\%.  
The $b$-tagging efficiency uncertainty of $5.3\%$ results in an uncertainty on the
cross sections with one $b$ jet of $5.3\%$ and on those with two $b$
jets of $10.6\%$. Additional uncertainties arise from the jet energy
scale~\cite{jesnim} ($2.4\%$) and the backgrounds as described in
section~\ref{sec:bgs} ($2.0\%$).

All these uncertainties apply to the ratio of the $Z+\bjet$ to the $Z$ cross section. 
For the $Z+\bjet$ cross section itself, 
additional uncertainties apply due to the uncertainty 
on the integrated luminosity (5.8\%)~\cite{lumerr} and on the CDF measurement
of the $Z$ cross section (1.8\%)~\cite{wzinclprl}. 

While the per-jet cross section is independent of the Monte Carlo
model used for the number of $\bjets$ in each event, the event based
cross sections depend on the assumption on this number. We estimate a
systematic uncertainty on the ratio of events with two $\bjets$ to one $\bjet$ 
of 30\% as determined from the measurement of the cross section ratio for one
and two
$\bjets$ presented in section~\ref{sec:results}. This results in an
additional uncertainty of up to 4.7\% on any event based cross
section.

For the measurement of ${\sigma^{\rm evt}(Z+2\bjets)}/{\sigma(Z)}$ 
an uncertainty of 100\% on the $c$ and light jet backgrounds is taken, 
resulting in an uncertainty of $12\%$ on the cross section ratio.

\section{Results}
\label{sec:results}
In this section the integrated and differential measurements for the $Z+\bjet$ cross section
divided by the inclusive $Z$ cross section are presented. Also shown are the integrated cross section
and the integrated cross section divided by the $Z$+jet cross section.

The measurements are compared to the leading-order QCD Monte Carlo
generators {\sc pythia} and {\sc alpgen}, and to the next-to-leading
order calculations as implemented in {\sc mcfm}~\cite{zbmcfm}. The QCD
calculations are always performed in the same kinematic range as the
data. 

The {\sc mcfm} calculation is performed at order $\alpha_s^2$. The 
$gb \to Zb$ is dependent on the $b$ quark density, and is
performed at next-to-leading order in $\alpha_s$. The other processes
contributing at order $\alpha_s^2$ are the final states $Zbg$ and $Zb\bar{b}$,
which are calculated at leading order. 
The $b$ quarks are treated as massless throughout, except in the
contribution $q\bar{q} \to Zb\bar{b}$ where the quark mass is 
required in order to render the calculation finite. The NLO 
corrections are known to substantially increase the cross section
for the $gb \to Zb$ process and to decrease its dependence on renormalization and
factorization scales. For the $q\bar{q} \to Zb\bar{b}$ process no full NLO calculations
are available for the case where only one $b$ jet is observed. This leads
to a substantial uncertainty on the cross section as discussed below.  
For the results presented here two predictions are compared: $Q^2=m_Z^2+p_{T,Z}^2$ and 
$Q^2=(\sum_{i=1}^{N_{\rm jet}} p_{T,i}^2)/N_{\rm jet}=\langle p_{T,{\rm jet}}^2\rangle$. The same scales
are used for the renormalization and factorization scale for the two predictions.

{\sc alpgen} is a tree-level generator where the partonic initial and final states are 
showered using {\sc pythia}. In the evaluation of the matrix elements, {\sc alpgen} treats 
$b$ quarks as massive, and therefore $b$ quarks cannot be considered as parts of the partonic 
density of the proton. The inclusive $Z+b$ final state emerges in {\sc alpgen} as part of the 
full $gg \to Z b \bar{b}$ process  after summing over the full phase-space of the $\bar{b}$ quark. 
The result is dominated by configurations where one of the initial-state gluons splits into a 
$b\bar{b}$ pair: here, the $b$ quark enters the hard scattering with the second gluon leading 
to the $Z+b$ final state, and the $\bar{b}$ typically has small transverse momentum and large rapidity.
For {\sc alpgen}  the default renormalization and factorization scales of $Q^2=m_Z^2+p_{T,Z}^2$ are used.

{\sc pythia} includes the $b\bar{b} \to Z$ process as part of the generic
$q\bar{q} \to Z$ process, with a cross section related to the
parton density of the incoming $b\bar{b}$ quarks evaluated at the
$Z$ mass scale. Then, by backwards evolution of the initial-state
cascade, two branchings $g \to b\bar{b}$ are constructed to promote the
original process to $gg \to Z b\bar{b}$, where the scale is set by the
transverse momentum (in some approximation) of each branching on
its own. In addition further partons may be emitted in the cascade,
and one may also have light quark $q\bar{q} \to Z$ processes where
final-state $g \to b\bar{b}$ branchings gives $Zb\bar{b}$ topologies.

For {\sc mcfm} the CTEQ6M~\cite{cteq6} parton distribution functions
are used, while for {\sc pythia} and {\sc alpgen} the CTEQ5L set is used. 

Corrections for the underlying
event and the hadronization (see section~\ref{sec:selection}) 
are determined using the {\sc pythia}
Monte Carlo and applied to the {\sc mcfm} prediction. The underlying
event correction is determined by taking the difference in the cross
section with and without the underlying event switched on. The
hadronization correction is determined by taking the difference in the
cross section at the parton and hadron level.  For ${\sigma^{\rm
jet}(Z+\bjet)}/{\sigma(Z)}$ and the integrated $Z+\bjet$ cross section, 
these corrections result in a net increase of the predicted cross
section by $8\%$ with contributions of $-1\%$ from hadronization and
$+9\%$ from the underlying event. For the cross section  ratio of $Z+\bjet$ to inclusive $Z+$jet production the  correction
is $+4\%$ with contributions of $+9\%$ from hadronization and
$-5\%$ from the underlying event. The correction factors for the differential cross section ratios are given below.

The ratio of the integrated $Z+\bjet$ cross section for
$E_T^{\bjet}>20$~GeV and $|\eta^{\bjet}|<1.5$ to inclusive $Z$
production, for $76<M_{ll}<106$~GeV/$c^2$, is measured as
$$\frac{\sigma^{\rm jet}(Z+\bjet)}{\sigma(Z)}=(3.32 \pm 0.53 {\rm
(stat.)} \pm 0.42 {\rm (syst.)}) \times 10^{-3}.$$ 
This measurement is proportional to the number of $b$ jets.
The NLO QCD prediction of {\sc mcfm} is $2.3 \times 10^{-3}$ for $Q^2=m_Z^2+p_{T,Z}^2$
and $2.8 \times 10^{-3}$ for $Q^2=\langle p_{T,{\rm jet}}^2\rangle$. 
The prediction of {\sc alpgen} is $2.1\times 10^{-3}$ and {\sc pythia} predicts $3.5\times 10^{-3}$.
The  difference between the two {\sc mcfm} predictions shows that there is a
rather large theoretical uncertainty for this process. The reason for the large difference
between {\sc alpgen} and {\sc pythia} is primarily due to the use of different
scales since {\sc alpgen} uses a large scale while {\sc pythia}'s scale is approximately
the jet $p_T$. The data are better described by a low choice of scale.

The ratio of $Z+\bjet$ to inclusive $Z+$jet production
is determined as $(2.08 \pm 0.33 \pm 0.34)\%$ compared to predictions
of $1.8\%$ ({\sc mcfm}, $Q^2=m_Z^2+p_{T,Z}^2$), $2.2 \%$ ({\sc mcfm}, 
$Q^2=\langle p_{T,{\rm jet}}^2\rangle$), $1.5\%$ ({\sc alpgen}), and $2.2\%$ ({\sc pythia}).
The $Z+\bjet$ cross section is determined to be $\sigma^{\rm
jet}(Z+\bjet)=0.85 \pm 0.14 {\rm (stat.)} \pm 0.12 {\rm (syst.)}$~pb
by multiplying the ratio with the measured inclusive $Z$ cross section
from CDF of $254.9 \pm 16.2$~pb~\cite{wzinclprl}.
This technique means  there is an implicit (small) extrapolation from the measurement 
range $76<M_{ll}<106$~GeV/$c^2$ presented in this article
to the region $66<M_{ll}<116$~GeV/$c^2$, where the inclusive $Z$ cross section
measurement was made.

Table~\ref{tab:xsec} gives the differential results for the ratio of
the $Z+\bjet$ cross section to the inclusive $Z$ production cross
section versus the $E_T^{\bjet}$ and $\eta^{\bjet}$.  These
measurements are proportional to the number of $b$ jets. 
Table~\ref{tab:xsev} lists the differential cross section ratios versus
$p_T^Z$, together with the  ratio for one and
two jets and for one and two $b$ jets. These measurements are
proportional to the number of events. Also included
in Tables~\ref{tab:xsec} and  ~\ref{tab:xsev}
is the correction factor $C_{\rm had}$ that needs to be
applied to parton level calculations to correct for the underlying
event and hadronization.

\begin{table}[htbp]
\begin{center}
\caption{
The ratio of the $Z+\bjet$ to the inclusive $Z$ cross section versus $E_T^{\bjet}$ 
(normalized per GeV) and $\eta^{\bjet}$ (normalized per unit in pseudorapidity).
The statistical uncertainty is listed first and the 
systematic uncertainty is listed second. The correction factor $C_{\rm had}$ that needs to be applied
to parton level calculations to correct for the underlying
event and hadronization is also given.\label{tab:xsec}}
\vspace{0.1cm}

\begin{tabular}{lcc} 
\hline \hline
$E_T^{\bjet}$ (GeV) & $\sigma^{\rm jet}(Z+\bjet)/\sigma(Z) \times 10^{4}$ (GeV)$^{-1}$ & $C_{\rm had}$\\\hline
$[20,35]$  & $1.42 \pm 0.28 \pm 0.15$    & $1.03$ \\
$[35,55]$  & $0.25 \pm 0.10 \pm 0.03$    & $1.13$ \\
$[55,100]$ & $0.122 \pm 0.043 \pm 0.019$ & $1.22$ \\\hline
$|\eta^{\bjet}|$ & $\sigma^{\rm jet}(Z+\bjet)/\sigma(Z) \times 10^{3}$ & $C_{\rm had}$\\\hline
$[0.0,0.5]$    & $2.44 \pm 0.57 \pm  0.28$ & $1.13$ \\
$[0.5,1.0]$    & $2.90 \pm 0.65 \pm  0.39$ & $1.03$ \\
$[1.0,1.5]$    & $0.79 \pm 0.50 \pm  0.14$ & $1.05$ \\
\hline \hline
\end{tabular}
\end{center}
\end{table}

\begin{table}[htbp]
\begin{center}
%\vspace{1mm}
\caption{
The ratio of the $Z+\bjet$ to the inclusive $Z$ cross section versus
the number of jets,  the number of $\bjets$, and the $p_T$ of the $Z$ boson 
for events with at least one $\bjet$.
In all cases the measurement is restricted to 
$E_T^{\bjet}>20$~GeV and $|\eta^{\bjet}|<1.5$. The statistical uncertainty is listed first and the 
systematic uncertainty is listed second. The correction factor $C_{\rm had}$ that needs to be applied
to parton level calculations to correct for the underlying
event and hadronization is also given. \label{tab:xsev}}
\begin{tabular}{lcc} 
\hline \hline
$p_T^Z$  (GeV/$c$) & $\sigma^{\rm evt}(Z+\geq 1 \bjet)/\sigma(Z) \times 10^{5}$ (GeV/$c$)$^{-1}$ & $C_{\rm had}$\\\hline
$[0,20]$ & $4.6 \pm 1.4 \pm 0.7$ & 1.25 \\
$[20,35]$& $7.0 \pm 1.9 \pm 0.8$ & 1.09 \\
$[35,55]$& $2.9 \pm 1.0 \pm 0.2$ & 0.93 \\
$[55,100]$& $1.11 \pm 0.39 \pm 0.15$ & 1.14 \\\hline

$N_{\rm jet}$ & $\sigma^{\rm evt}(Z+\geq 1 \bjet)/\sigma(Z) \times 10^{3}$ & $C_{\rm had}$\\\hline
$1$         & $2.23 \pm 0.42 \pm 0.28$ & 1.07 \\
$2$         & $0.78 \pm 0.22 \pm 0.09$ & 1.12\\\hline
$N_{\bjet}$ & $\sigma^{\rm evt}(Z+ N_{\bjet})/\sigma(Z) \times 10^{3}$ & $C_{\rm had}$\\\hline
$1$     & $2.75 \pm 0.44 \pm 0.38$ & 1.07 \\
$2$     & $0.22 \pm 0.11 \pm 0.05$ & 1.09\\\hline\hline
\end{tabular}
\end{center}
\end{table}

\begin{figure}[htbp]
  \begin{center}
    \includegraphics*[width=\figurewidth]{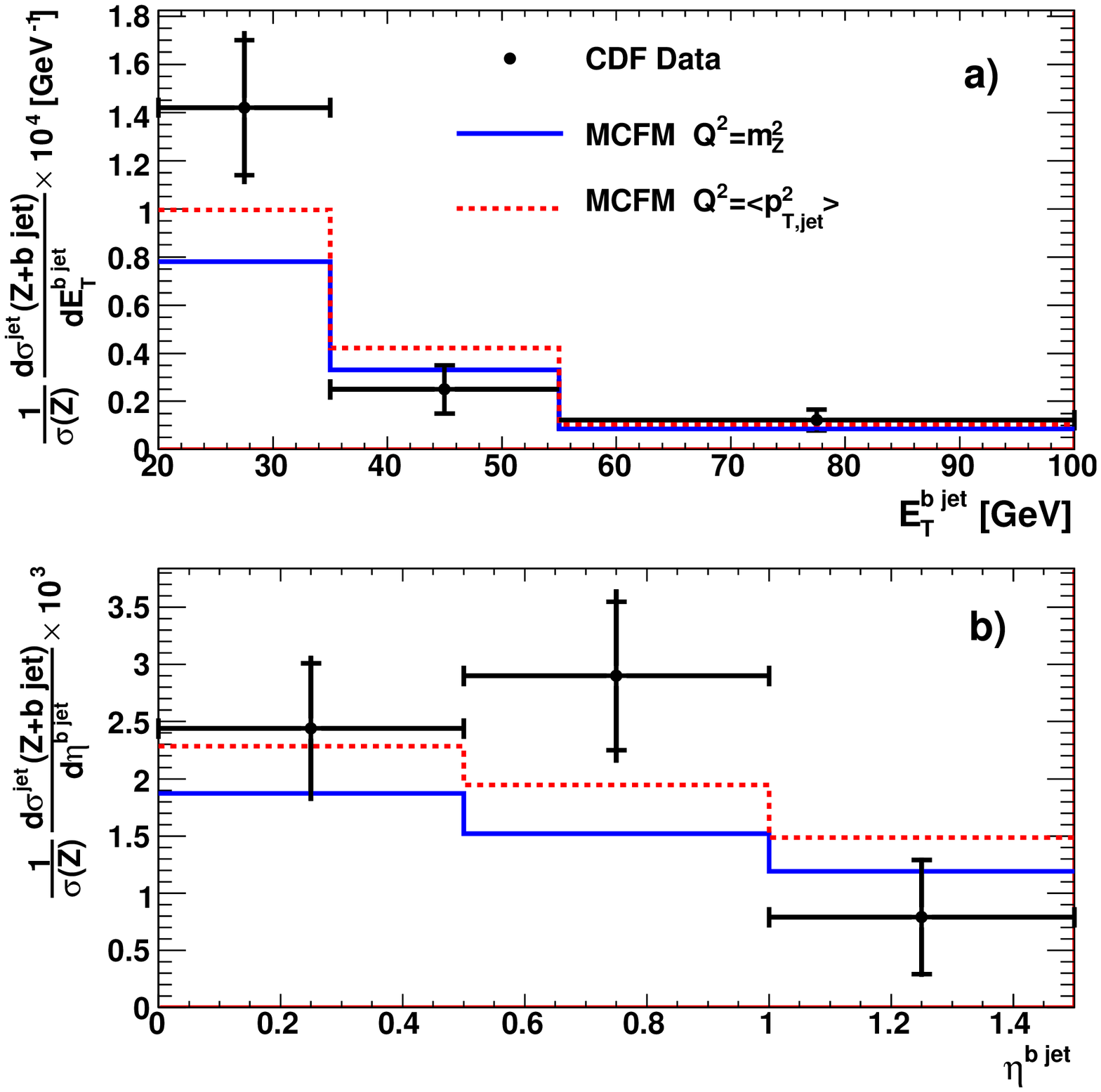}
  \end{center}

    \caption{Ratio of the $Z+\bjet$ to the inclusive $Z$ boson cross
    section versus a) $E_T^{\bjet}$ and b) $\eta^{\bjet}$. Shown are
    the data (points) compared to the predictions from {\sc mcfm}
    calculated with the scales $Q^2=m_Z^2+p_{T,Z}^2$ (solid line) and
    with $Q^2=\langle p_{T,{\rm jet}}^2\rangle$ (dotted line). The
    inner error bars represent the statistical errors, and the outer
    error bars represent the total errors.
\label{fig:dsdptetamcfm}}
\end{figure}

\begin{figure}[htbp]
  \begin{center}
    \includegraphics*[width=\figurewidth]{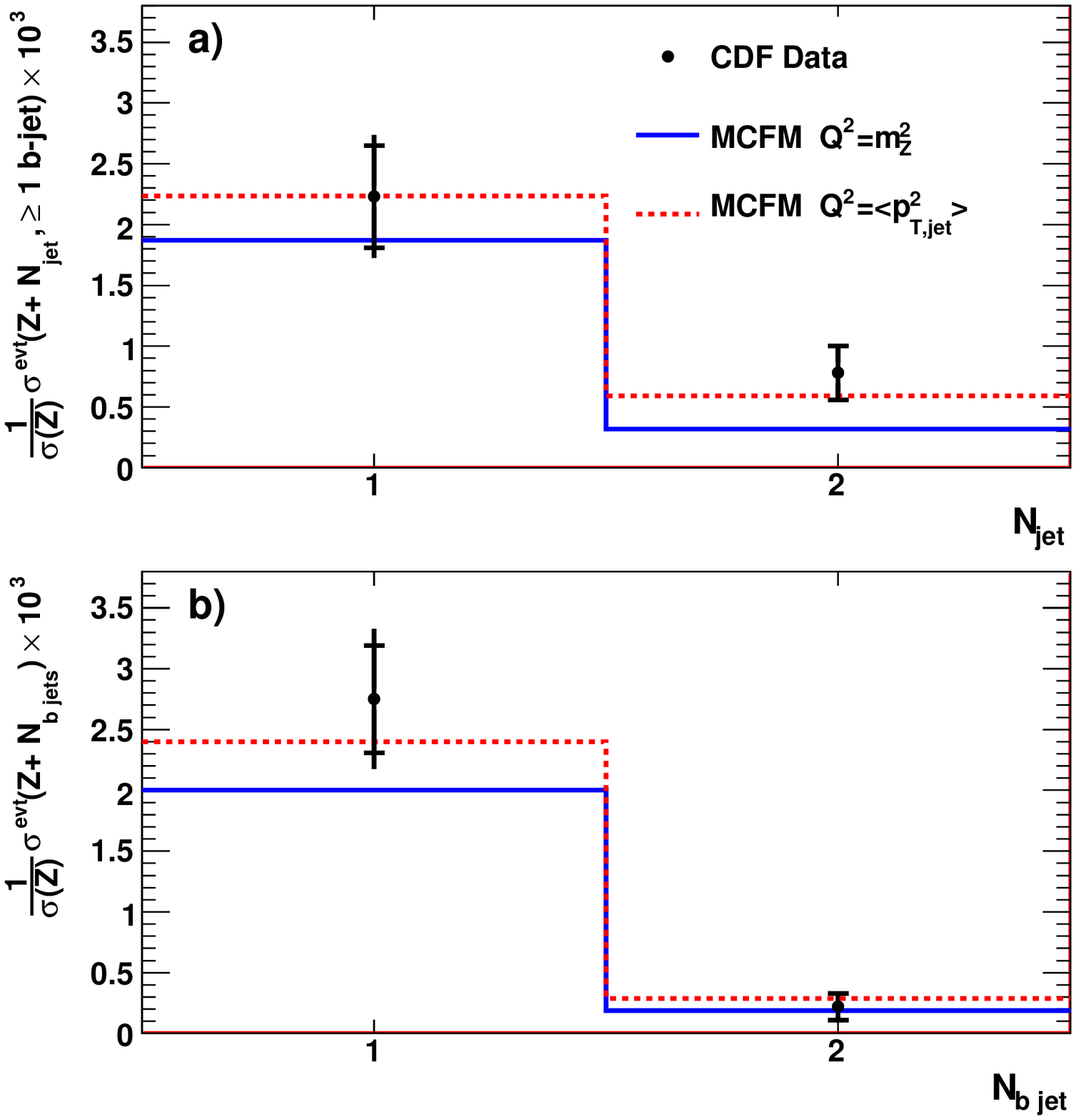}
  \end{center}
    \caption{Ratio of the $Z+\bjet$ cross section to the 
$Z$ boson cross section versus a) $N_{\rm jet}$ and 
b) $N_{\bjet}$ .Shown are
    the data (points) compared to the predictions from {\sc mcfm}
    calculated with the scales $Q^2=m_Z^2+p_{T,Z}^2$ (solid line) and
    with $Q^2=\langle p_{T,{\rm jet}}^2\rangle$ (dotted line). The
    inner error bars represent the statistical errors, and the outer
    error bars represent the total errors.
\label{fig:dsdnjetmcfm}}
\end{figure}

\begin{figure}[htbp]
  \begin{center}
    \includegraphics*[width=\figurewidth]{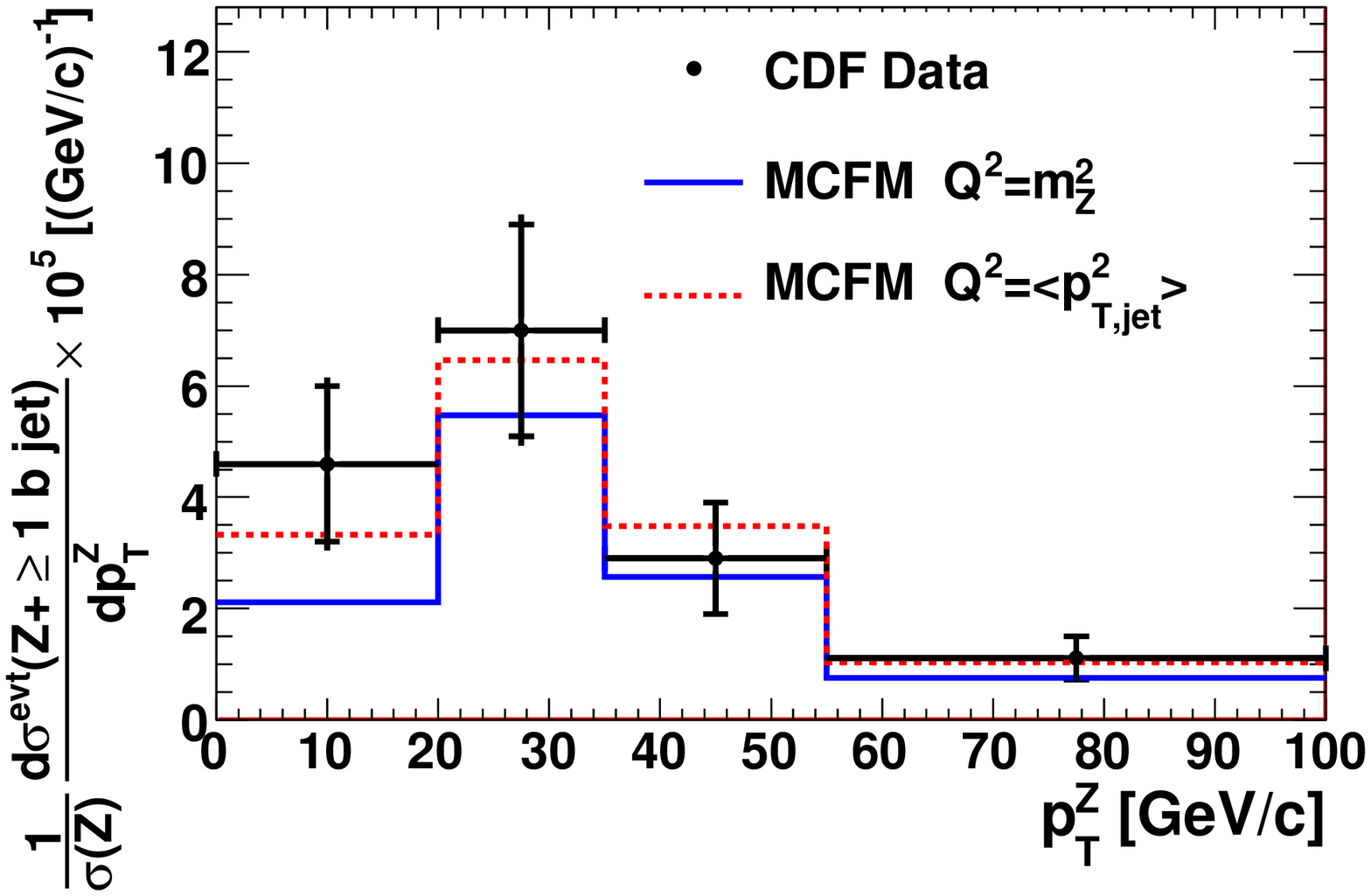}
  \end{center}
    \caption{Ratio of the $Z+\bjet$ cross section to the 
$Z$ boson cross section versus  $p_T(Z)$. Shown are
    the data (points) compared to the predictions from {\sc mcfm}
    calculated with the scales $Q^2=m_Z^2+p_{T,Z}^2$ (solid line) and
    $Q^2=\langle p_{T,{\rm jet}}^2\rangle$ (dotted line). The
    inner error bars represent the statistical errors, and the outer
    error bars represent the total errors.
\label{fig:dsdptzmcfm}}
\end{figure}

Figs.~\ref{fig:dsdptetamcfm}, \ref{fig:dsdnjetmcfm}, and
\ref{fig:dsdptzmcfm} show the data compared to the {\sc mcfm} 
 prediction versus $E_T^{\bjet}$ and $\eta^{\bjet}$,
versus the number of jets and $b$ jets, and versus $p^Z_T$,
respectively. The {\sc mcfm} predictions are shown for two different
values for the renormalization and factorization scale.

It is seen that the theoretical cross section prediction depends on the choice
of scale, and differences up to a factor of two are seen, e.g., in the $N_{\rm jet}$ distribution.
Both predictions describe the data but the lower scale choice is favored.

\begin{figure}[htbp]
  \begin{center}
    \includegraphics*[width=\figurewidth]{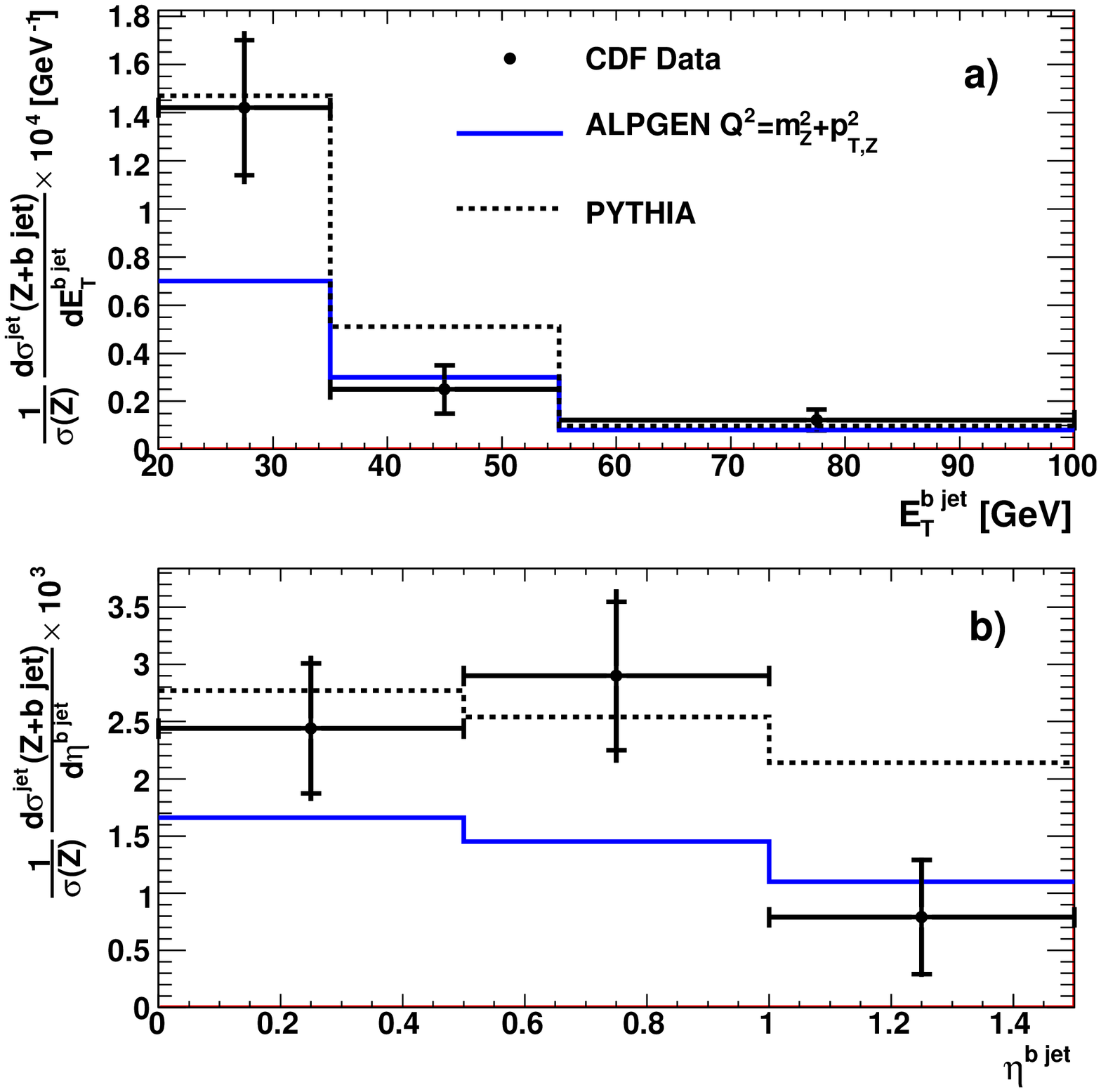}
  \end{center}
    \caption{Ratio of the $Z+\bjet$  to the inclusive $Z$ boson cross section 
versus a) $E_T^{\rm jet}$ and b) $\eta^{\rm jet}$. Shown are the data (points) compared
to the predictions from  {\sc alpgen}   (solid line)  and {\sc pythia} (dotted line). The inner error bars represent the statistical errors, and the outer error bars represent the total errors.
\label{fig:dsdptetamc}}
\end{figure}

\begin{figure}[htbp]
  \begin{center}
    \includegraphics*[width=\figurewidth]{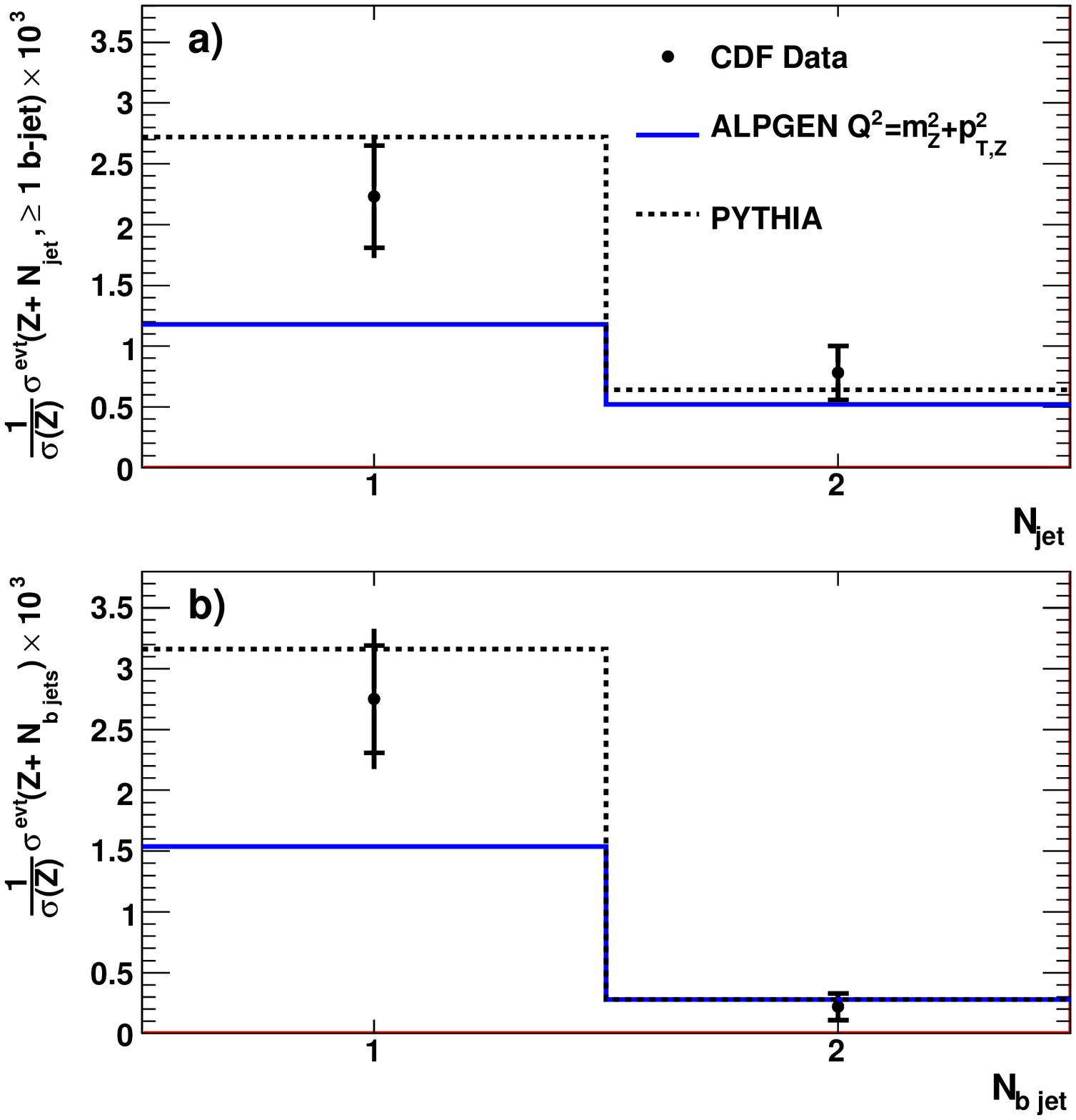}
  \end{center}
    \caption{Ratio of the $Z+\bjet$ cross section to the 
$Z$ boson cross section versus a) $N_{\rm jet}$ and 
b) $N_{\bjet}$.  Shown are the data (points) compared
to the predictions from  {\sc alpgen}   (solid line)  and {\sc pythia} (dotted line). The inner error bars represent the statistical errors, and the outer error bars represent the total errors.
%with scales $Q^2=m^2_Z+p^2_{T,Z}$ (dashed line)
% and $Q^2=\sum m^2_b+p^2_{T,b}$
\label{fig:dsdnjetmc}}
\end{figure}

\begin{figure}[htbp]
  \begin{center}
    \includegraphics*[width=\figurewidth]{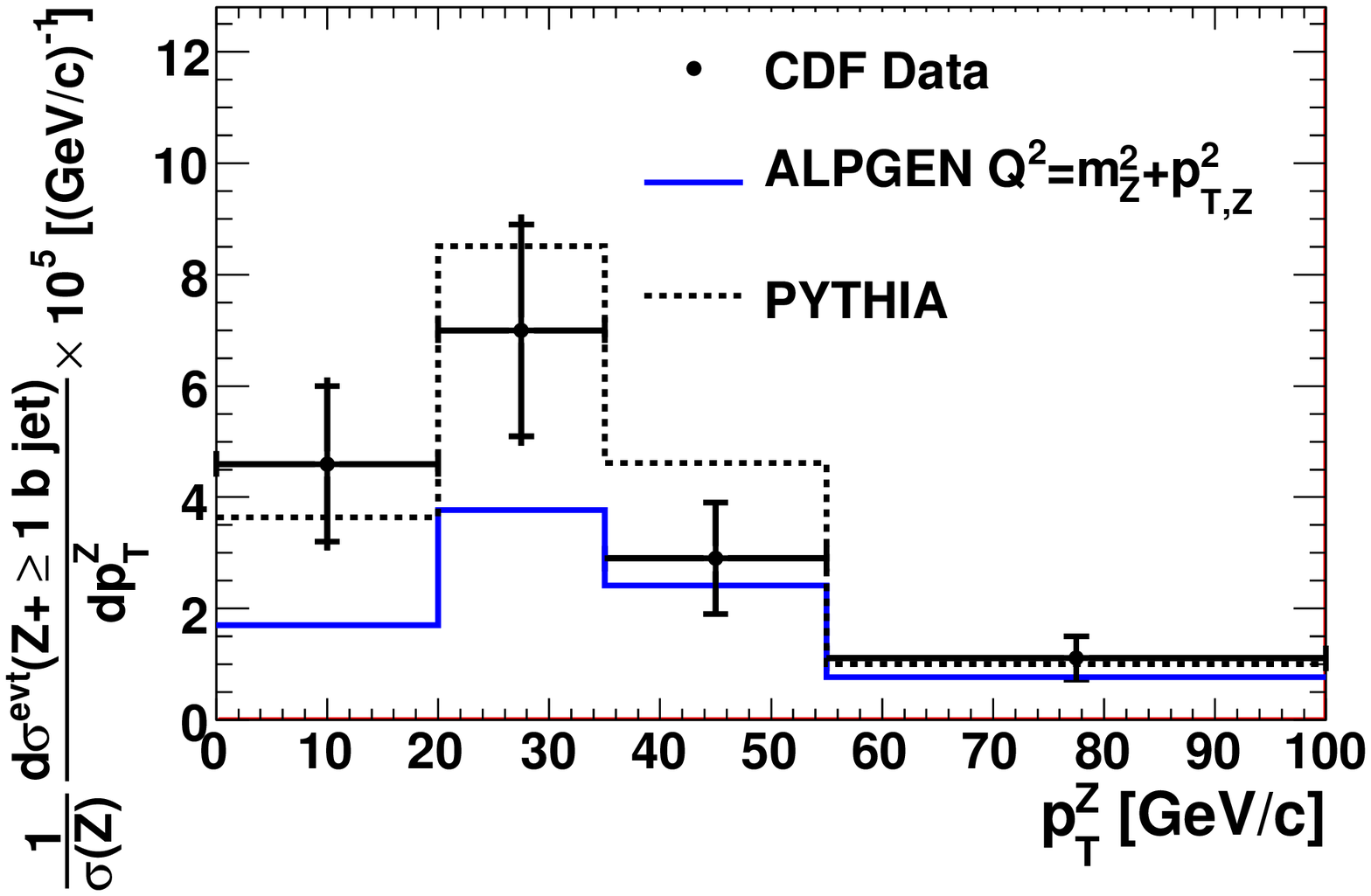}
  \end{center}
    \caption{Ratio of the $Z+\bjet$ cross section to the 
$Z$ boson cross section versus  $p_T(Z)$. Shown are the data (points) compared
to the predictions from  {\sc alpgen}   (solid line)  and {\sc pythia} (dotted line). The inner error bars represent the statistical errors, and the outer error bars represent the total errors.
\label{fig:dsdptzmc}}
\end{figure}

Figs.~\ref{fig:dsdptetamc}, \ref{fig:dsdnjetmc}, and \ref{fig:dsdptzmc}
show the data compared to the {\sc alpgen} and {\sc pythia} Monte
Carlo programs. Large differences are observed between the two
programs, in particular at low $E_T^{\bjet}$ and low $p_T^Z$ and at
low jet multiplicity. We have verified that this difference is reduced
if we use a lower scale for {\sc alpgen} but present here only the
default used commonly by hadron collider experiments. In general {\sc
pythia} describes the data better than {\sc alpgen}. Both MC programs
describe the data well at high $E_T^{\bjet}$ and $p_T^Z$ and for jet
multiplicities of two.

All predictions are  generally in agreement  with the data,  but
differences of up to $2\sigma$ are observed in the integrated cross
section between the data and the {\sc mcfm} calculation, depending on which scale
is used. The large spread of the theoretical predictions
suggests that higher orders in the QCD calculation may be 
important for this process.

%In particular the {\sc mcfm} prediction for the process $q\bar{q} \to Z b\bar{b}$ where
%either both $b$ quarks are inside the same $b$ jet or one of the two $b$ jets is not
%within the kinematic range $E_T^{\bjet}>20$~GeV and $|\eta^{\bjet}|<1.5$ 
%is only available at leading order and has a large scale uncertainty. Since this process
%contributes about half the total cross section this results in a large uncertainty on the 
%$Z+\bjet$ production rate.

\section{Conclusion}
In conclusion, we have measured the ratio of $Z+\bjet$ production to
inclusive $Z$ production using the CDF II detector at the Tevatron, and 
compared to previous
measurements the uncertainty has been reduced
to 20\%. For the first time we have presented 
differential measurements as a function of the kinematics of the jets
and $Z$ boson and the number of jets in the event. These
measurements enable the NLO QCD prediction to be tested over a wide range
of final state observables. Large variations are seen between the theoretical
predictions as no full NLO QCD calculation is available for this process.
The predictions generally describe the data, but the agreement is better 
for those predictions that use a low value for the renormalization and
factorization scales.

%, although the NLO QCD prediction is seen to be about $2\sigma$
%lower than the data. The largest deviation is seen at low
%$E_T^{\bjet}$. 

\section*{Acknowledgments}
We thank the Fermilab staff and the technical staffs of the
participating institutions for their vital contributions. We are
thankful to J. Campbell, F. Maltoni, M. Mangano, M. Seymour, T.
Sj\"ostrand and J. Thaler for the many interesting and helpful
discussions regarding the theoretical predictions. This work was
supported by the U.S. Department of Energy and National Science
Foundation; the Italian Istituto Nazionale di Fisica Nucleare; the
Ministry of Education, Culture, Sports, Science and Technology of
Japan; the Natural Sciences and Engineering Research Council of
Canada; the National Science Council of the Republic of China; the
Swiss National Science Foundation; the A.P. Sloan Foundation; the
Bundesministerium f\"ur Bildung und Forschung, Germany; the Korean
Science and Engineering Foundation and the Korean Research Foundation;
the Science and Technology Facilities Council and the Royal Society,
UK; the Institut National de Physique Nucleaire et Physique des
Particules/CNRS; the Russian Foundation for Basic Research; the
Ministerio de Ciencia e Innovaci\'{o}n, and Programa
Consolider-Ingenio 2010, Spain; the Slovak R\&D Agency; and the
Academy of Finland.

%

%\bibitem{btageff} ksome reference for b-tagging efficiency


\begin{thebibliography}{99} 
% list_of_visitor_addresses_r1.tex                            8/31/01
%

\bibitem{zbmcfm} J. Campbell, R.K. Ellis, F. Maltoni, and S. Willenbrock, Phys. Rev. D~{\bf 69}, 074021 (2004).

\bibitem{pythia} 
T.~Sj\"{o}strand, L. L\"{o}nnblad, and S.~Mrenna, LU-TP 01, arXiv:hep-ph/0108264.
%http:\/\/www.thep.lu.se/ $\tilde{ }$ torbjorn/.

\bibitem{alpgen} M. Mangano, M. Moretti, and R. Pittau, Nucl. Phys. B~{\bf 632}, 343  (2002);

M. Mangano~{\it et al.},  J. High Energy Phys.~{\bf 0307}, 001 (2003).

\bibitem{smhiggs} M. Carena {\it et al.}, (Higgs Working Group Collaboration), arXiv:hep-ph/0010338.

\bibitem{sbottomd0} V. M. Abazov {\it et al.} (D0 Collaboration), Phys. Rev. Lett.~{\bf 97}, 171806 (2006).

\bibitem{sbottomcdf} T. Aaltonen {\it et al.} (CDF Collaboration), Phys. Rev. D~{\bf 76}, 072010 (2007). 

\bibitem{singletop} T. Stelzer, Z. Sullivan, and S. Willenbrock, Phys. Rev. D~{\bf 56}, 5919 (1997);

T. Stelzer, Z. Sullivan, and S. Willenbrock, Phys. Rev. D~{\bf 58}, 094021 (1998).


\bibitem{bbh}
D. Dicus, T. Stelzer, Z. Sullivan, and S. Willenbrock, Phys. Rev. D~{\bf 59}, 094016 (1999);

C. Balazs, H. J. He, and C. P. Yuan, Phys. Rev. D~{\bf 60}, 114001 (1999);

F. Maltoni, Z. Sullivan, and S. Willenbrock, Phys. Rev. D~{\bf 67}, 093005 (2003).



\bibitem{mssmhiggs} D. Choudhury, A. Datta, and S. Raychaudhuri, arXiv:hep-ph/9809552; 

C. S. Huang and S. H. Zhu, Phys. Rev. D~{\bf 60}, 075012 (1999);

J. Campbell, R.K. Ellis, F. Maltoni, and S. Willenbrock, Phys. Rev. D~{\bf 67}, 095002 (2003).




\bibitem{cdfzb} A. Abulencia {\it et al.} (CDF Collaboration), Phys. Rev. D~{\bf 74}, 032009 (2006).

\bibitem{d0zb}  V. M. Abazov {\it et al.} (D0 Collaboration), Phys. Rev. Lett.~{\bf 94}, 161801 (2005).




\bibitem{cdfzjet} T. Aaltonen {\it et al.} (CDF Collaboration), Phys. Rev. Lett.~{\bf 100}, 102001 (2008).

\bibitem{d0zjet} V. M. Abazov {\it et al.} (D0 Collaboration), Phys. Lett. B~{\bf 658}, 112 (2008). 

\bibitem{Wb} C. Neu (for the CDF Collaboration), {\it $W^\pm/ Z$ + Jets and $W^\pm$ / Z + Heavy Flavor Jets at the Tevatron.}, Proceedings of 43rd Rencontres de Moriond on QCD 
and
Hadronic Interactions, La Thuile, Italy, 2008.


\bibitem{wzinclprl} 
A. Abulencia~\textit{et al.} (CDF Collaboration), J.~Phys.~G~{\bf 34}, 2457 (2007).

\bibitem{CDF_run2} D. Acosta \textit{et al.} (CDF Collaboration), 
Phys. Rev. D~{\bf 71}, 032001 (2005).


\bibitem{silicon} C. S. Hill \textit{et al}., Nucl. Instrum. Methods A~{\bf 530}, 1 (2004);

A. Sill \textit{et al}., Nucl. Instrum. Methods A~{\bf 447}, 1 (2000);

A. Affolder \textit{et al}., Nucl. Instrum. Methods A~{\bf 453}, 84 (2000).

\bibitem{COT} A. Affolder \textit{et al}., Nucl. Instrum. Methods A~{\bf 526}, 249 (2004).

\bibitem{cem} L. Balka \textit{et al.}, Nucl. Instrum. Methods A~{\bf 267}, 272 (1988).

\bibitem{pem}  M. Albrow \textit{et al.}, Nucl. Instrum. Methods A~{\bf 480}, 524 (2002).

\bibitem{cha}  S. Bertolucci \textit{et al.}, Nucl. Instrum. Methods A~{\bf 267}, 301 (1988).

\bibitem{ces} S. Kuhlmann {\it et al.}, Nucl. Instrum. Methods  A {\bf 518}, 39 (2004).

\bibitem{cmu}  G. Ascoli \textit{et al.}, Nucl. Instrum. Methods A~{\bf 268}, 33 (1988).

%\bibitem{cdfjpsi}  D. Acosta {\it et al.} (CDF Collaboration), Phys. Rev. D~{\bf 71}, 032001 (2005).

\bibitem{CLC} D. Acosta  \textit{et al.}, Nucl. Instrum. Methods A~{\bf 461}, 540 (2001).

\bibitem{pythiaz} G. Miu and T. Sj\"{o}strand, Phys. Lett. B~{\bf 449}, 313 (1999).

\bibitem{mlm} M. Mangano~{\it et al.},  J. High Energy Phys.~{\bf 0701}, 013 (2007).

\bibitem{cteq5l} 
H.L.~Lai \textit{et al}., Eur. Phys. J. C~{\bf 12}, 375 (2000).

\bibitem{ue} 
D. Acosta \textit{et al.} (CDF  Collaboration), Phys. Rev. D~{\bf 70}, 072002 (2004).
\bibitem{rickdwt} R. D. Field (for the CDF Collaboration), 
{\it Studying the Underlying Event at CDF}, Proceedings of 33rd
International Conference on High Energy Physics (ICHEP 06), Moscow, Russia, 2006.

\bibitem{evtgen} D.~J.~Lange, Nucl. Instrum. Methods A~{\bf 462}, 152 (2001).

\bibitem{geant} GEANT, Detector description and simulation tool,
CERN Program Library Long Writeup W5013 (1993).

\bibitem{cdfsim} 
E.~Gerchtein and M.~Paulini,
 {\it CDF detector simulation framework and performance},
Proceedings of 2003 Conference for Computing in High-Energy and Nuclear
 Physics (CHEP 03), La Jolla, California, 2003.

%E. Gerchtein and M. Paulini, Computing in High Energy and 
%Nuclear Physics,  arXiv:physics-0306031, (2003).

\bibitem{wchargeasym} D. Acosta {\it et al.} (CDF Collaboration), Phys. Rev. D~{\bf 71}, 051103 (2005). 

\bibitem{cone}  G. Arnison  \textit{et al}., Phys. Lett. B~{\bf 123}, 115 (1983).

\bibitem{jesnim} A. Bhatti \textit{et al}., Nucl. Instrum. Methods A~{\bf 566}, 375 (2006).

\bibitem{secvtx} 
%% top cross section paper
D. Acosta \textit{et al.} (CDF Collaboration), Phys. Rev. D~{\bf 71}, 052002 (2005).


\bibitem{zz} J. Campbell and R.K. Ellis, Phys. Rev. D~{\bf 60}, 113006 (1999). 

\bibitem{ttbar} N. Kidonakis and R. Vogt, Eur. Phys. J. C~{\bf 33}, 466 (2004).

\bibitem{trilprd} T. Aaltonen \textit{et al}., (The CDF Collaboration), Phys. Rev. D~{\bf 77}, 052002 (2008).

\bibitem{lumerr} S. Klimenko, J. Konigsberg, and T.M. Liss, FERMILAB-FN-0741 (2003).


\bibitem{cteq6} 
J. Pumplin {\it et al.},
%``New Generation of Parton Distributions with Uncertainties from Global QCD Analysis.''
J. High Energy Phys.~{\bf 0207}, 012 (2002);


%\bibitem{collins} M. A. Aivazis, J. C. Collins, F. I. Olness, and W. K. Tung, Phys. Rev. D~{\bf 50}, 3102 (1994); J. C. Collins, Phys. Rev. D~{\bf 58}, 094002 (1998); J. Pumplin  {\it et al.}, J. High Energy Phys.~{\bf 0207}, 012 (2002).

%\bibitem{h1} A. Aktas {\it et al.} (H1 Collaboration), Eur. Phys. J. C~{\bf 45},  23 (2006);
%A. Aktas \textit{et al}., Eur. Phys. J. C~{\bf 40},  349 (2005).
%\bibitem{cdfbplus}  D. Acosta {\it et al.} (CDF Collaboration), Phys. Rev. D~{\bf 65}, 052005 (2002).
%\bibitem{cdfjpsi}  D. Acosta {\it et al.} (CDF Collaboration), Phys. Rev. D~{\bf 71}, 032001 (2005).

%\bibitem{d0bjet} B. Abbott {\it et al.} (D0 Collaboration) , Phys. Rev. Lett.~{\bf 85}, 5068 (2000).

%\bibitem{pythia} Torbjorn Sj¨ostrand, Leif Lonnblad, and Stephen Mrenna. 
%Pythia 6.203: Physics and manual and http://www.thep.lu.se/torbjorn/pythia.html. 
%Comput. Phys. Commun., 135(238), 2001.




%\bibitem{puzzle} private communication with M. Mangano, F. Maltoni, S. Mrenna, J. Thaler
% NNLO Z cross section
%\bibitem{znnlo}
%P. Sutton, A. D. Martin, R. Roberts, and W. Stirling, Phys. Rev. D~{\bf 45}, 2349 (1992); 
%R. Rijken and W. van Neerven, Phys. Rev. D~{\bf 51}, 44 (1995); 
%R. Hamberg, W. van Neerven and T. Matsuura, Nucl. Phys. B~{\bf 359}, 343 (1991);
%R. Harlander and W. Kilgore, Phys. Rev. Lett.~{\bf 88}, 201801 (2002);
%W. van Neerven and E. Zijstra, Nucl. Phys. B~{\bf 382}, 11 (1992).

% ????? definition of HT
%\bibitem{ht} F. Abe \textit{et al.} (CDF Collaboration), Phys. Rev. Lett.~{\bf 75}, 3997 (1995).




%\bibitem{msecvtx}
%  D.~J.~Jackson,
%``A topological vertex reconstruction algorithm for hadronic jets,''
%  Nucl.\ Instrum.\ Methods\ A~{\bf 388} (1997) 247.
%  J.~Abdallah {\it et al.} (DELPHI Collaboration),
%``b-tagging in DELPHI at LEP,''
%  Eur.\ Phys.\ J.\ C~{\bf 32} (2004) 185.

% thesis from Monica
%\bibitem{monica} M. D'Onofrio, Ph.\ D.\ thesis, University of Geneva (2006).

%\bibitem{herwig}
%G. Corcella {\it et al.}, J. High Energy Phys.~{\bf 0101}, 010 (2001).

% b-quark fragmentation paper
%\bibitem{lepjetmulti}
%LEP/SLD Heavy Flavour Working Group, D.~Abbaneo {\it et al.},  
% LEPHF 2001-01;
%(available from
%http://lepewwg.web.cern.ch/LEPEWWG/heavy/).

\end{thebibliography}
\end{document}